\documentclass[prd,showpacs,preprintnumbers,groupedaddress,superscriptaddress,nofootinbib,11pt]{revtex4-1} 

\usepackage{amsthm,amsmath}
\newtheoremstyle{newplain}
{}
{}
{\it}
{}
{\bfseries}
{.}
{5pt}
{\thmname{#1}\hspace{5pt}\thmnumber{#2}\thmnote{\hspace{2pt}[{\normalfont #3}]}}
\theoremstyle{newplain}

\newtheorem{definition}{Definition}

\usepackage{cases}
\usepackage{ulem}
\usepackage{amsfonts}
\usepackage{mathrsfs}

\usepackage[pdftex]{graphicx}
\usepackage[hang,small,bf]{caption}
\usepackage[subrefformat=parens]{subcaption}
\usepackage{afterpage}

\usepackage{hyperref} 
\hypersetup{%
 setpagesize=false,
 bookmarksnumbered=true,%
 bookmarksopen=true,%
 colorlinks=true,%
 linkcolor=blue,%
 citecolor=red}

\usepackage{color}
\usepackage{comment}

\newcommand{\blue}[1]{\textcolor{blue}{#1}}
\newcommand{\mr}[1]{\mathrm{#1}}

\begin{document}

\preprint{OCU-PHYS-556}
\preprint{AP-GR-177}
\preprint{RUP-22-2}

\title{Dynamical photon sphere and time evolving shadow around black holes with temporal accretion}



\author{Yasutaka Koga}
\affiliation{Division of Particle and Astrophysical Science, Graduate School of Science, Nagoya University, Nagoya 464-8602, Japan}

\author{Nobuyuki Asaka}
\affiliation{Department of Mathematics and Physics, Osaka City University, Osaka 558-8585, Japan}

\author{Masashi Kimura}
\affiliation{Department of Physics, Rikkyo University, Toshima, Tokyo 171-8501, Japan}
\affiliation{Department of Information and Electronics, Daiichi Institute of Technology, Tokyo 110-0005, Japan}

\author{Kazumasa Okabayashi}
\affiliation{Department of Mathematics and Physics, Osaka City University, Osaka 558-8585, Japan}



\date{\today}

\begin{abstract}
A photon sphere is known as the geometrical structure that shapes a black hole shadow.
The mechanism is well understood for static or stationary black hole spacetimes such as the Schwarzschild and the Kerr spacetimes.
In this paper, we investigate and explicitly specify a photon sphere that shapes a black hole shadow in a dynamical spacetime while taking the global structure of the spacetime into account.
We consider dynamical and eternal black hole cases of the Vaidya spacetime, which represents a spherically symmetric black hole with accreting null dust.
First, we numerically show that there are the dynamical photon sphere and photon orbits corresponding to the shadow edge in a moderate accretion case.
Second, the photon spheres are derived analytically in special cases.
Finally, we discuss the relation between our photon sphere and the several notions defined as a photon sphere generalization.
\end{abstract}

\pacs{04.20.-q, 04.70.-s, 04.70.Bw}

\maketitle


\tableofcontents

\section{Introduction}
\label{sec:introduction}
A photon sphere, the radius of circular photon orbits, is known to play a key role in observation of a black hole shadow, for example, in the Schwarzschild spacetime.
Concerning black hole shadow observations, there are two important aspects.
First, the radius is a threshold for photons coming from distant light sources to escape to infinity or fall into the black hole.
In the observer's sight, there exists a region from which the photons cannot come in principle.
This dark region is called a black hole shadow.
Second, the photon sphere accumulates photons along the radius.
If light sources emit photons for a long time, an enormous number of photons orbit around the sphere and eventually escape to infinity.
Then the observer observes a very bright shadow edge corresponding to the photon sphere as actually observed by the Event Horizon Telescope~\cite{eht}.
\par
In static and spherically symmetric cases, we can analyze the escaping photon behaviors and what we will observe around a black hole by use of the conserved quantities, energy and angular momentum.
In the Schwarzschild case, the first analysis was given by Synge~\cite{synge}.
Pande and Durgapal gave the analysis in generic spherically symmetric configuration~\cite{Pande_1986}.
From the analyses, we can see that the radius of circular photon orbits, namely the photon sphere, is important for black hole shadow formation.
\par
In dynamical cases, it is challenging to define a photon sphere as a structure that shapes a black hole shadow even in spherical symmetry.
This is because there are not so many exact solutions to the Einstein equation that are physically reasonable and the geodesic equation does not reduces to one-dimensional radial potential problem due to the absence of the conserved energy.
Although several generalized notions of photon sphere have been proposed from various points of view~\cite{claudel,siino_2019,siino_2021,yoshino_tts,yoshino_dtts,kobialko_2020}, not so many examples in dynamical cases are known yet.
The aim of this paper is to specify dynamical photon spheres that shape black hole shadows in specific cases.
\par
One of good models for this problem is the Vaidya spacetime, an exact solution to the Einstein equation with accreting null dust~\cite{Vaidya_1951}.
The spacetime metric looks like the Schwarzschild spacetime in Eddington-Finkelstein coordinates $(v,r,\theta,\phi)$ with the mass $M$ replaced by the arbitrary mass function $m(v)$.
We can model an accreting black hole and gravitational collapse by specifying the mass function appropriately.
As a preceding work, Mishra, Chakraborty, and Sarkar~\cite{Mishra_2019} investigated photon spheres of the Vaidya spacetime with several mass functions and showed their evolution in the future characteristic time regions.
Solanki and Perlick~\cite{Solanki_2022} investigated the Vaidya spacetime by assuming a linearly increasing mass functions over the entire time region and specified the photon sphere analytically by using the self-similarity of the spacetime.
See also Ref.~\cite{Sarkar_2021} for works on variations of the Vaidya spacetime.
\par
In this paper, we investigate photon spheres and black hole shadows of the Vaidya spacetime focusing on the first aspect of the importance mentioned above.
That is, we suppose that black hole shadows are formed not due to the red shift of photons, but due to the absence of null geodesics that emanate from a distant light source and reach the corresponding points on the celestial sky of a distant observer.
We assume the mass function to be exactly constant in the past and future so that the global structure becomes as simple as the Schwarzschild spacetime and define the photon sphere from the causal point of view.
As our main focus, we show the evolution of the photon sphere for the entire time region and clarify what the appropriate boundary condition is.
Specifying the photon sphere as analytically as possible, we also discuss the relation between our photon sphere and several generalized notions of a photon sphere.
\par
In Sec.~\ref{sec:review}, reviewing the photon sphere and the black hole shadow of the Schwarzschild spacetime, we clarify what is the photon sphere shaping a black hole shadow in the Vaidya spacetime.
In the current work, we focus our attention on eternal black holes that are static in the past and future time domain.
We specify the photon spheres and the behaviors of null geodesic motions corresponding to the edge of the black hole shadows in the following three cases.
In Sec.~\ref{sec:case1}, we consider the case where the black hole increases its mass moderately and numerically show the photon sphere shaping the black hole shadow.
The result briefly shows the existence of a photon sphere relevant to a black hole shadow in a dynamical case.
In Sec.~\ref{sec:analytical}, we consider linearly increasing mass in the dynamical time domain.
The photon sphere is explicitly specified in terms of the parameters of the spacetime, such as the initial and final mass.
In Sec.~\ref{sec:analytical-shell}, the null dust shell accretion is considered.
In Sec.~\ref{sec:discussion}, we discuss the relation between our photon sphere and recently proposed generalized notions of a photon sphere.
In Sec.~\ref{sec:summary}, we summarize our results.
We use units in which $G=1$ and $c=1$. 

\afterpage{\clearpage}
\newpage

\section{Review and Preliminary}
\label{sec:review}
Reviewing the photon sphere and the black hole shadow in the Schwarzschild spacetime, we clarify what we investigate as a photon sphere and a black hole shadow in the Vaidya spacetime.
\subsection{Schwarzschild photon sphere and the black hole shadow}
\begin{figure}[h]
\centering
\includegraphics[width=200pt]{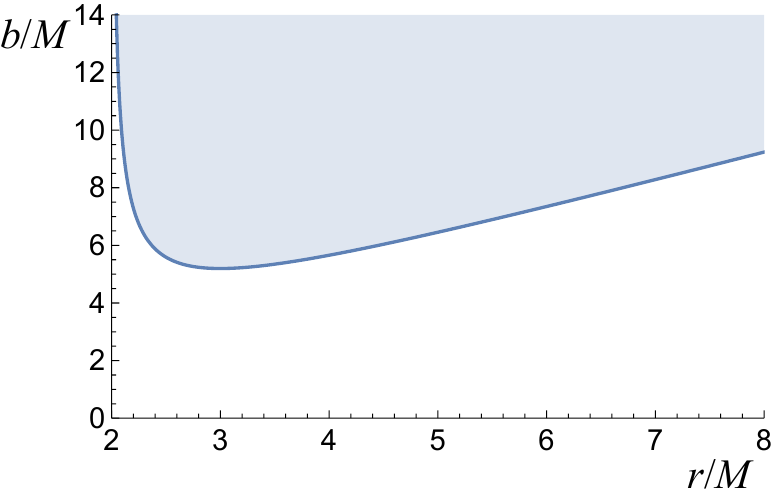}
\caption{
\label{fig:schwarzschild-potential}
$r$-$b$ plane for null geodesics in the Schwarzschild spacetime.
The region of $V(b;r)>0$ (shaded region) is the forbidden region.
Null geodesics with $b>b_\text{c}$ coming from infinity are eventually reflected by the potential at radii corresponding to $V(b;r)=0$.
}
\end{figure}

The spacetime is given by the metric
\begin{equation}
g=-\left(1-\frac{2M}{r}\right)dt^2+\left(1-\frac{2M}{r}\right)^{-1}dr^2+r^2d\Omega_2^2.
\end{equation}
A null geodesic obeys the one-dimensional potential problem,
\begin{equation}
\dot{r}^2+V(E,L;r)=0,\;\; V(E,L;r):=L^2\left(1-\frac{2M}{r}\right)r^{-2}-E^2,
\end{equation}
where $\dot{}:=d/d\lambda$ is the derivative by the affine parameter $\lambda$, $E:=-g(\partial_t,k)$ and $L:=g(\partial_\phi,k)$ are the energy and the angular momentum, and $k$ is the null geodesic tangent.
We have assumed that the null geodesics are in the equatorial plane $\theta=\pi/2$, without loss of generality.
Normalizing the affine parameter as $\lambda\to\lambda/E$, the equation reduces to
\begin{equation}
\dot{r}^2+V(b;r)=0,\;\; V(b;r):=V(1,b;r),
\label{eq:effpotential2}
\end{equation}
where $b:=L/E$ is the impact parameter.
Null geodesics are drawn as horizontal lines in $r$-$b$ plane (Fig.~\ref{fig:schwarzschild-potential}).
They are reflected by the effective potential when they touch the boundary of the forbidden region, $V(b;r)>0$.
The extremum of $V(b;r)$ is at $r=3M$.
This is the Schwarzschild photon sphere.
\par
\begin{figure}[t]
    \includegraphics[width=100pt]{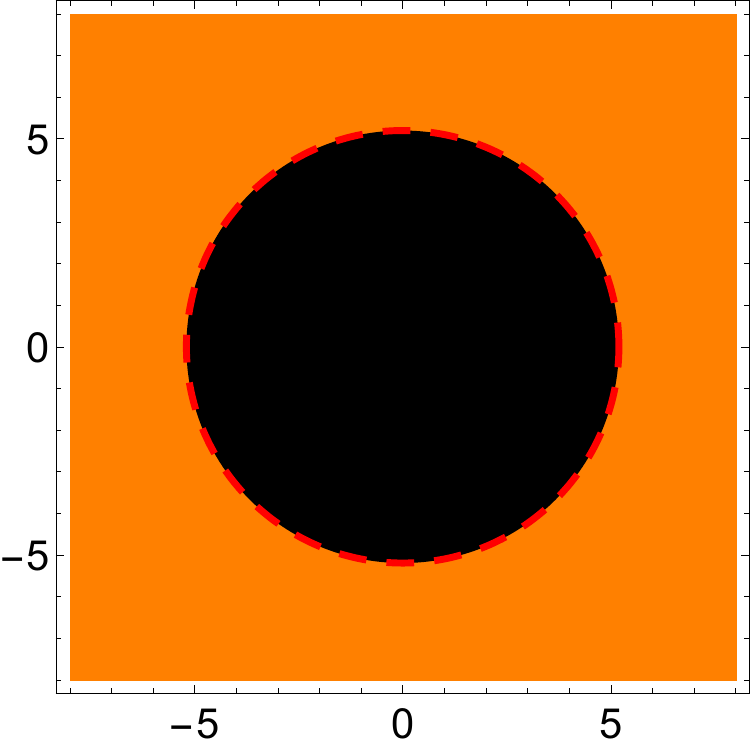}
    \caption{Image of the black hole shadow in the Schwarzschild spacetime. The bright and dark regions in the observer's sight are shown as the orange and black regions. The distance from the center corresponds to the impact parameter, and the shadow edge (red dashed line) corresponds to the critical impact parameter.
    }
    \label{fig:Shadowimage_Sch}
\end{figure}

Suppose an observer looking to the black hole at $r_{\mr{obs}}\gg2M$ and a spherical light source at $r_{\mr{src}}>r_{\mr{obs}}$.
There is the critical impact parameter, $b_\text{c}=3\sqrt{3}M \simeq 5.196 M$.
For $|b|>b_\text{c}$, null geodesics emanating from the source are reflected by the potential and eventually reach $r_{\mr{obs}}$.
For $|b|<b_\text{c}$, null geodesics from the source fall into the black hole.
For the null geodesics reaching $r_{\mr{obs}}$, the incident angle $\alpha$ to the observer is given by
\begin{equation}
k=\beta(e_0+\cos\alpha e_1+\sin\alpha e_2),
\end{equation}
where $\beta$ is a constant and the tetrad $\{e_\mu\}$ for the equatorial plane is given by
\begin{equation}
e_0=\sqrt{1-\frac{2M}{r}}^{-1}\partial_t,\;\;
e_1=\sqrt{1-\frac{2M}{r}}\partial_r,\;\;
e_2=r^{-1}\partial_\phi.
\end{equation}
Using the impact parameter, we have
\begin{equation}
\tan\alpha = \frac{g(k,e_2)}{g(k,e_1)}
=\frac{b\sqrt{\left(1-2M/r_{\mr{obs}}\right)r_{\mr{obs}}^{-2}}}{\sqrt{1-b^2\left(1-2M/r_{\mr{obs}}\right)r_{\mr{obs}}^{-2}}}
\end{equation}
or, by approximation with the large $r_{\mr{obs}}$,
\begin{equation}
\label{eq:shadow-size}
\alpha\simeq\frac{b}{r_{\mr{obs}}}.
\end{equation}
Since $|b|>b_\text{c}$ for null geodesics reaching the observer, the least impact parameter $b_\text{c}$ determines the apparent angular size of the dark region, i.e., the black hole shadow, as $\alpha_{\mr{sh}}=b_\text{c}/r_{\mr{obs}}$.\footnote{
If we take $\{X,Y\}$ as Cartesian coordinates of the observer's celestial sky with the origin corresponding to the line of sight to the black hole, then the incident angle corresponds to the radius, $\sqrt{X^2+Y^2}=|\alpha|$.
The normalization, $X\to X/r_{\mr{obs}}, Y\to Y/r_{\mr{obs}}$, gives the relation, $\sqrt{X^2+Y^2}=|b|$.
In this sense, the shadow size is often said to be $b_\text{c}$.}
Furthermore, since the near critical null geodesics with $|b|=b_\text{c}+0$ are the orbits asymptoting $r=3M$, the photon sphere is said to shape the black hole shadow.
Note that the formula, Eq.~\eqref{eq:shadow-size}, is valid for other asymptotically flat spacetimes.
Figure~\ref{fig:Shadowimage_Sch} shows the image of the black hole shadow for a distant observer surrounded by a spherical light source.
Even in dynamical black hole spacetimes, the impact parameter of the marginally escaping null geodesics determines the shadow size.

\begin{figure}[t]
    \includegraphics[width=200pt]{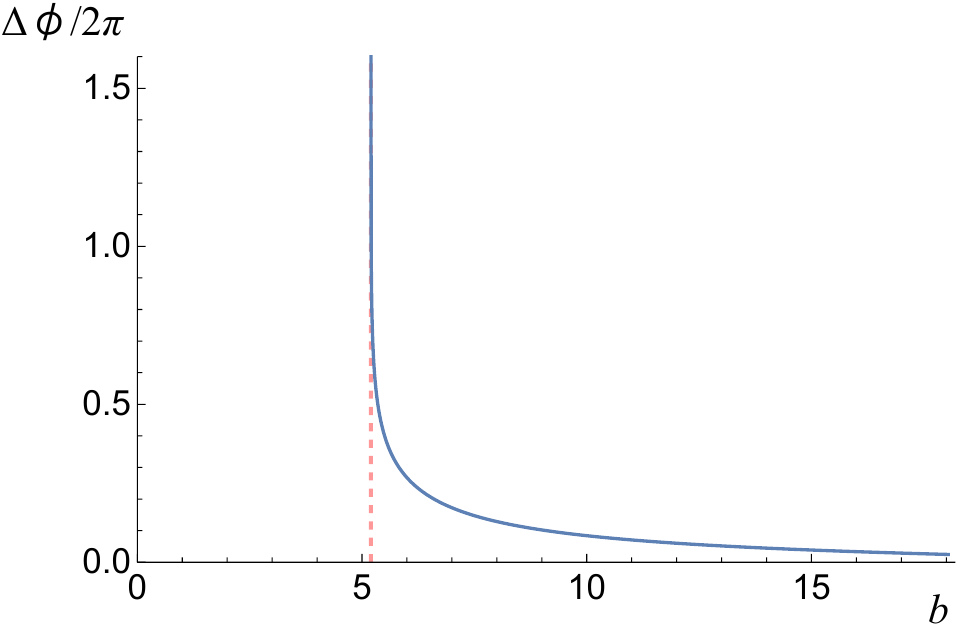}
    \caption{Winding number of null geodesics in the Schwarzschild spacetime as a function of the impact parameter $b$. 
    The red dashed line denotes the critical impact parameter $b = b_\text{c} = 3 \sqrt{3}M \simeq 5.196 M$.
    We set $M = 1$ in the numerical calculation.
    }
    \label{fig:deltaphi_Sch}
\end{figure}

We consider a null geodesic which is emitted from a distant point and
reflected at the turning point whose radius is $r_{\rm min}$,
and reaches a distant observer.
From $L = r^2 d\phi/d\lambda$ and Eq.~\eqref{eq:effpotential2}, we obtain
\begin{align}
r^2 \frac{d\phi}{dr} = \pm \left(b^{-2} - r^{-2}\left(1-\frac{2M}{r}\right)\right)^{-1/2}.
\end{align}
Integrating this equation, we can define the winding number as
$n = \Delta \phi/(2\pi)$, where $\Delta \phi$ is
\begin{align}
\Delta \phi
=
2 \int_0^{1/r_{\rm min}} \frac{du}{\sqrt{b^{-2} - u^2(1- 2 M u)}} -\pi,
\end{align}
and $u := 1/r$.
We can regard $\Delta \phi$ as a function of the impact parameter $b$.
In Fig.~\ref{fig:deltaphi_Sch}, the winding number is plotted and
it is divergent at $b = b_{c}$.
It is known that this divergent behavior is logarithmic $n \sim -\ln(b-b_\text{c})$~\cite{1959RSPSA.249..180D, Luminet:1979nyg, Bozza:2002zj}.

\subsection{Schwarzschild photon sphere from the causal point of view}
From the causal point of view, the Schwarzschild photon sphere can be characterized as a hypersurface generated by null geodesics from $i^-$ to $i^+$.
This fact is important for the photon sphere to be a structure that shapes a black hole shadow for the following reason.
\par
\begin{figure}[t]
\includegraphics[width=200pt]{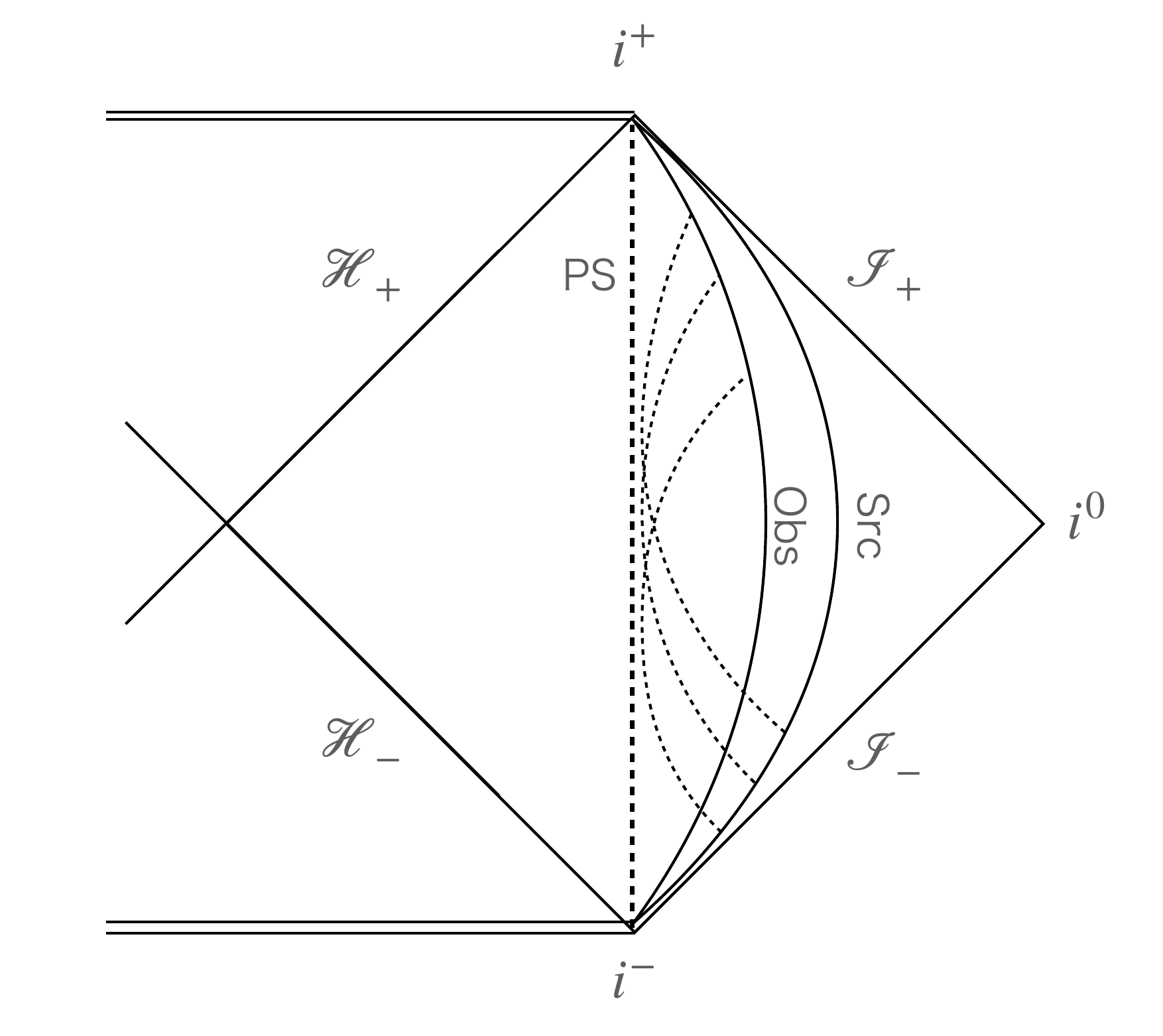}
\caption{\label{fig:shadow-setup-gralla}
The photon sphere (PS) and null geodesics (dashed lines) from a distant spherical light source to a distant observer (solid lines from $i^-$ to $i^+$) in the Schwarzschild spacetime.
}
\end{figure}

As a simple setup for a black hole shadow observation, one may suppose a distant observer looking to the black hole at $r=r_{\mr{obs}}\gg2M$ and a distant light source filling a sphere of $r=r_{\mr{src}}>r_{\mr{obs}}$~\cite{Cunha:2018acu, Gralla_2019, Okabayashi_2020}.
They are described as timelike curves from $i^-$ to $i^+$ in the Penrose diagram as in Fig.~\ref{fig:shadow-setup-gralla}.
What the observer observes is understood by the behavior of the past-directed null geodesics from each point of the observer's world line.
That is, the orbits of observed photons correspond to the null geodesics connecting the observer's and the source's world lines, and mapping of their impact parameters to the observer's celestial sky gives the shadow image.
Note that photons are supposed to be observed if they enter the observer's sight from the front, i.e., photons are outgoing when observed.
\par
\begin{figure}[h]
\includegraphics[width=200pt]{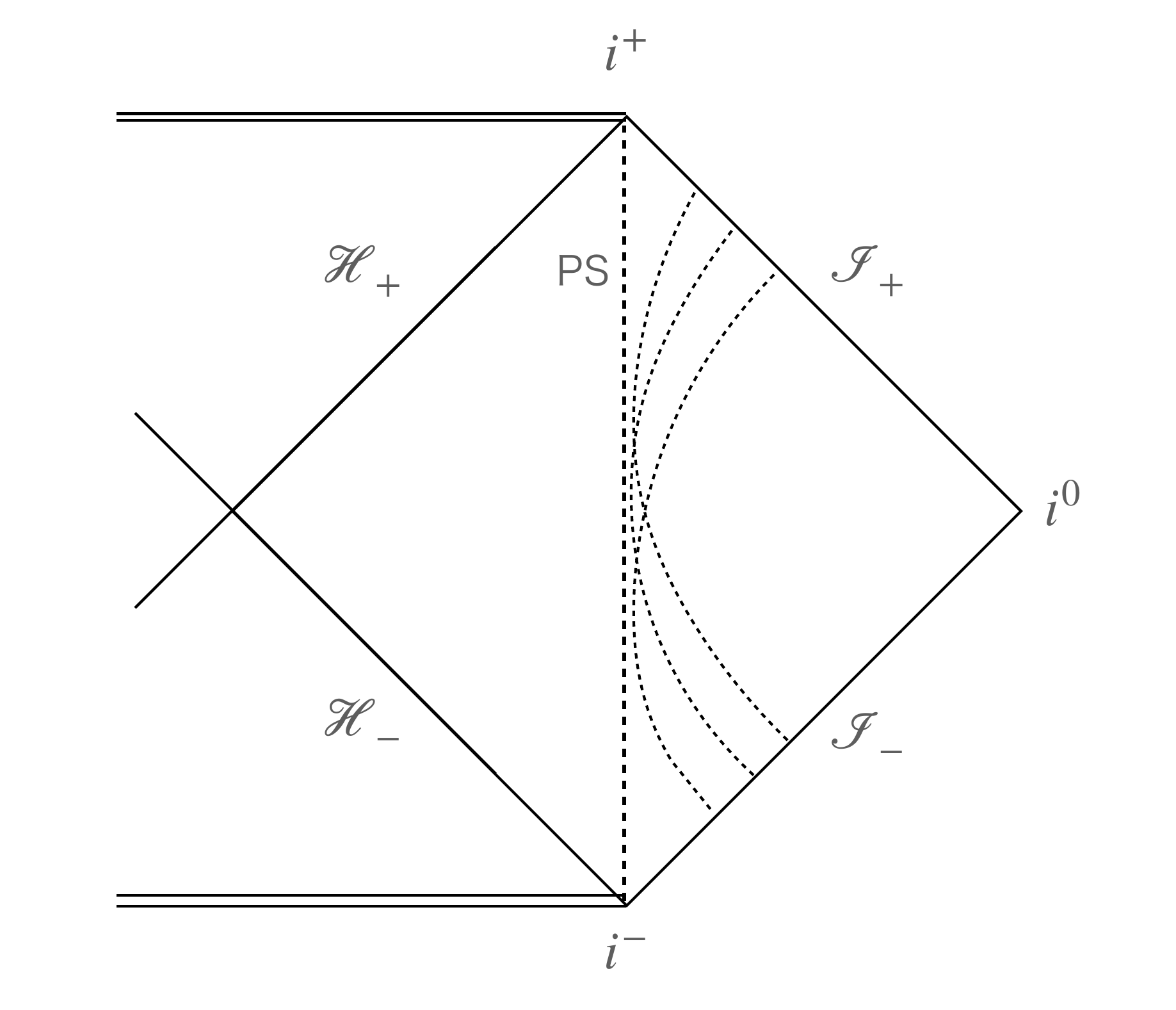}
\caption{\label{fig:shadow-setup-causal}
The PS and null geodesics (dashed lines) from the idealized light source, $\mathscr{I}^-$, to the idealized observer $\mathscr{I}^+$.
}
\end{figure}

Here we further idealize the situation from the causal point of view by taking the limit $r_{\mr{src}},\;r_{\mr{obs}}\to \infty$.
As we are concerned with null geodesics from the light source to the observer, the observer are supposed to be in the future of the light source, and therefore, we identify $\mathscr{I}^+$ and $\mathscr{I}^-$ with the idealized observer and source, respectively.
Note that, from the assumption that photons are outgoing when observed, we can ignore the case where the light source is on $\mathscr{I}^+$.
The past-directed null geodesics from $\mathscr{I}^+$ are classified into two types, ones to $\mathscr{I}^-$ and ones to $\mathscr{H}^-$.
\footnote{
Strictly speaking, there is only one null geodesic going to $i^-$ for each point on $\mathscr{I}^+$, ignoring those that can be identified by rotation reduced from the spherical symmetry of the spacetime.
Such a null geodesic has the exact critical impact parameter $b=b_\text{c}$.
}
The photon sphere then works as the boundary of the set of null geodesics from $\mathscr{I}^+$ and $\mathscr{I}^-$ (Fig.~\ref{fig:shadow-setup-causal}).
In particular, the null geodesics corresponding to the shadow edge (i.e., ones with $b=b_\text{c}+0$ in terms of the impact parameter) are the orbits that asymptote to the photon sphere.
Therefore, the causal feature of the photon sphere is actually important for the black hole shadow formation.

\subsection{Photon sphere of dynamical eternal black hole}
\begin{figure}[h]
\includegraphics[width=400pt]{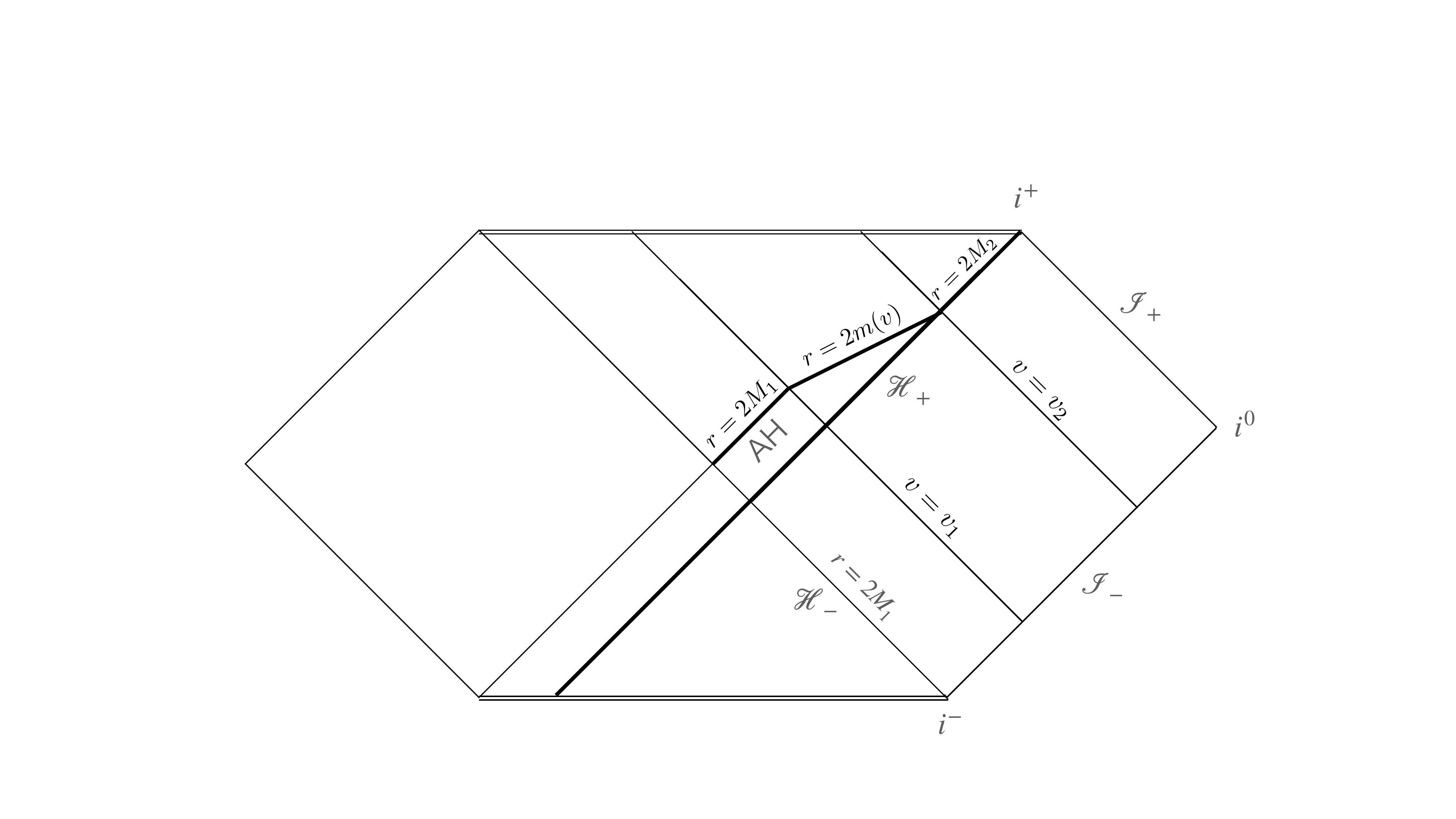}
\caption{\label{fig:eternal-vaidya}
The causal structure of the Vaidya spacetime with temporal accretion for $v_1<v\le v_2$.
The apparent horizon (AH) is given by $r=2m(v)$~\cite{nielsen} and does not coincide with the future event horizon $\mathscr{H}_+$ for $v\le v_2$.
}
\end{figure}

Even in a dynamical spacetime, the above causal argument for the photon sphere would hold if the causal structure is the same and the geometrical structure is not so different.
In this paper, we consider the Vaidya spacetime~\cite{Vaidya_1951},
\begin{equation}
g=-f(v,r)dv^2+2dvdr+r^2d\Omega^2,\;\; f(v,r)=1-\frac{2m(v)}{r},
\label{eq:vaidyametric}
\end{equation}
where the mass function $m(v)$ is arbitrary.
The Vaidya spacetime is the spherically symmetric black hole solution to the Einstein equation with null dust accretion.
By setting the mass function appropriately, we consider dynamical and eternal black hole cases.
We assume that the mass function $m(v)$ is initially constant [$m(v)=M_1$ for $v\le v_1$], monotonically increases temporally for $v_1<v\le v_2$, and finally constant [$m(v)=M_2$ for $v>v_2$].
The configuration guarantees the null energy condition and avoids the appearance of a naked singularity and the absence of future null infinity~\cite{Hiscock_1982,Kuroda_1984}.
Outside the horizons $\mathscr{H}_\pm$, the spacetime has the same causal structure as the Schwarzschild spacetime as shown in Fig.~\ref{fig:eternal-vaidya}.
\par
We investigate the photon sphere of the dynamical black hole spacetime especially focusing on its property as a structure that shapes the black hole shadow.
We define the dynamical photon sphere as a hypersurface generated by null geodesics from $i^-$ to $i^+$.
Trivial examples for such hypersurfaces are also shown in Appendix~\ref{sec:conf-sch}, however, we do not investigate them in detail because the causal structures are out of our scope.
Concerning the black hole shadow, we define {\it shadow edge orbits} as follows.
For each point on $\mathscr{I}^+$, consider the set of all the past-directed null geodesics emanating from the point and going to $\mathscr{I}^-$.
Among the set, we call the null geodesic having the smallest impact parameter at the point on $\mathscr{I}^+$ {\it a shadow edge orbit}.
\footnote{
This definition is thanks to the spherical symmetry and asymptotic flatness.
If the spacetime asymptotes to the Kerr metric in the far region for example, then we should define the shadow edge orbit in terms of the impact parameter associated with the Killing vector of rotation and the Carter constant associated with the Killing tensor.
}
If every shadow edge orbit asymptotes to the photon sphere, we can say that the photon sphere is the structure shaping the black hole shadow.
Note that this behavior of shadow edge orbits around the photon sphere implies that the photon sphere corresponds to {\it an unstable photon sphere} of a static case rather than {\it a stable photon sphere}, or {\it an anti-photon sphere}.
Although a stable photon sphere may also go from $i^-$ to $i^+$~\cite{gibbons_2016}, we focus on the dynamical photon sphere corresponding to an unstable photon sphere. 
\footnote{Since spacetimes of black holes formed by gravitational collapse have a quite different causal structure, it is not trivial whether our consideration and expectation for the dynamical photon spheres and shadow edge orbits are valid.}

When the spacetime is dynamical, it is generally difficult to discuss the radial geodesic equation in an analytic way, and it is often solved numerically. 
Hence, in the dynamical spacetime, finding the geodesic that goes to the timelike infinity is a difficult problem in general. 
However, in our case, since the spacetime in the past and future is the Schwarzschild spacetime, we just need to find a geodesic that has the critical impact parameter $b=b_{\mr{c}1}:=3\sqrt{3}M_1$ at $v=v_1$ and $b=b_{\mr{c}2}:=3\sqrt{3}M_2$ at $v=v_2$. Hence, we can find the geodesic from $i^-$ to $i^+$, i.e., the dynamical photon sphere with arbitrary accuracy by using the shooting method for the dynamical region.
By contrast, shadow edge orbits are obtained as follows. We numerically solve the null geodesic equation in the past direction from each time $v$ and some large $r$ for various impact parameters. The solutions going to $r\to \infty$ for $v\to -\infty$ are the photons observed by the observer, and among them, one that has the smallest impact parameter is the shadow edge orbit for each $v$. Then we see that the shadow edge orbits asymptote to the photon sphere.

\afterpage{\clearpage}
\newpage

\section{Numerical investigation of Vaidya photon sphere}
\label{sec:case1}
\begin{figure}[h]
\includegraphics[width=200pt]{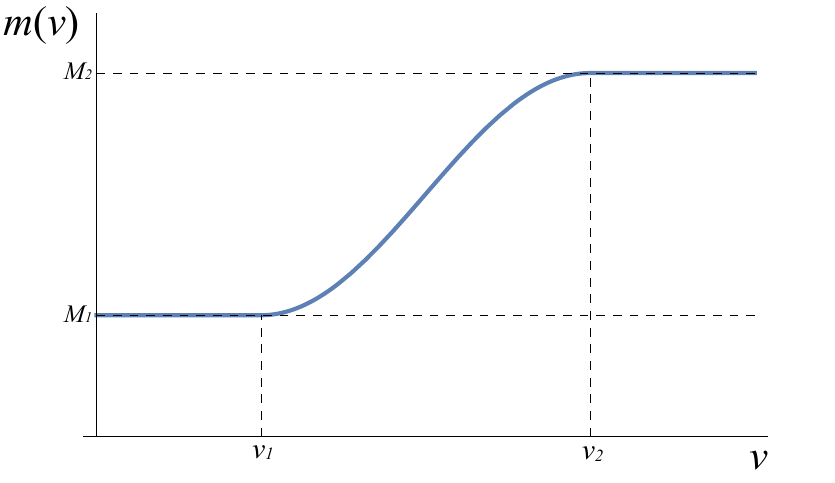}
\caption{\label{fig:mass-cos} 
The mass function~\eqref{eq:massfunction-cos} increases monotonically from $M_1$ to $M_2$.
}
\end{figure}

First we give an example of the photon sphere and shadow edge orbits by solving the geodesic equation numerically.
We consider the Vaidya spacetime, Eq.~\eqref{eq:vaidyametric}, with the mass function,
\begin{equation}
\label{eq:massfunction-cos}
m(v)=\left\{
\begin{array}{cc}
M_1 & v\le v_1\\
M_1+\dfrac{1}{2}(M_2-M_1)\left(1-\cos\left(\dfrac{v-v_1}{v_2-v_1}\pi\right)\right) & v_1<v\le v_2\\
M_2 & v>v_2
\end{array}
\right. .
\end{equation}
The spacetime is isometric to Schwarzschild spacetime with the mass $M_1$ and $M_2>M_1$ in the past time domain $v<v_1$ and the future time domain $v\ge v_2$, respectively.
In the intermediate dynamical domain, the mass increases monotonically.
The mass function moderately increases and is of class $C^1$ as shown in Fig.~\ref{fig:mass-cos}.

The photon sphere generator is a null geodesic with finite radius in far past and future.
In the current case, it implies that the null geodesic must asymptote to or coincide with $r=3M_1$ in the past and $r=3M_2$ in the future.
This is the boundary condition for the photon sphere generator.

\begin{figure}[h]
\includegraphics[width=200pt]{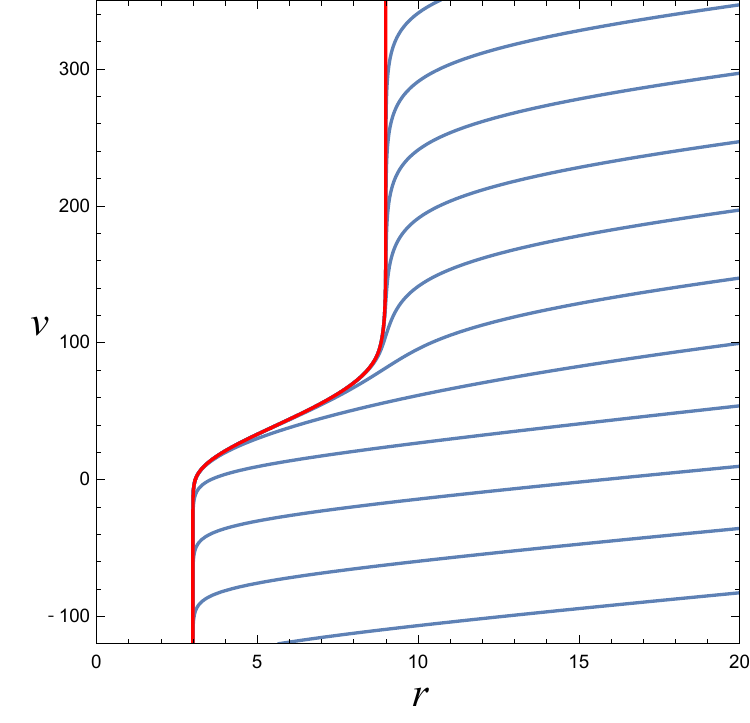}
\caption{\label{fig:psgenerator-numerical} 
The orbits of the photon sphere generator (red line) and shadow edge orbits that once approach the generator (blue line)
in the Vaidya spacetime with the mass function~\eqref{eq:massfunction-cos}. We set the parameters as $M_1=1$, $M_2=3$, $v_1=0$, and $v_2=100$.
The photon sphere generator asymptotes to $r=3M_1+0$ in the past and $r=3M_2-0$ in the future.
}
\end{figure}

\begin{figure}[h]
\includegraphics[width=180pt]{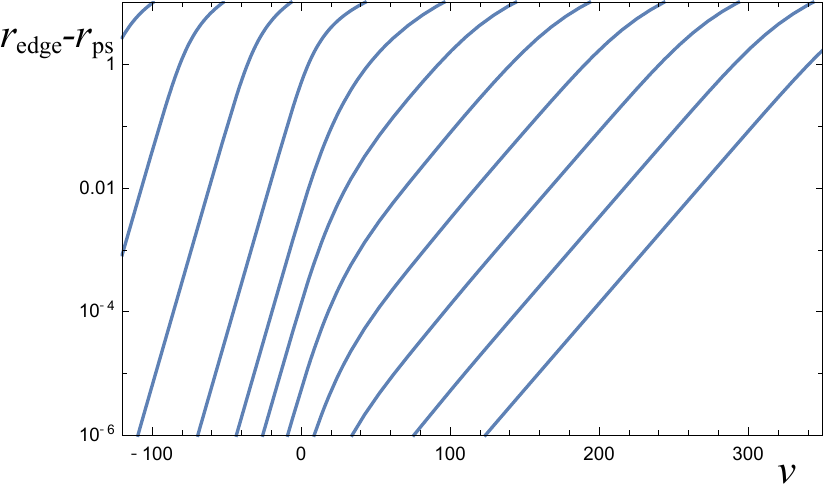}
\caption{\label{fig:psgenerator-numerical-logplot} 
The shadow edge orbits radius subtracted by the photon sphere generator radius.
We took the same parameters as in Fig.~\ref{fig:psgenerator-numerical}.
This shows that the shadow edge orbits 
asymptote to the photon sphere generator exponentially in time.
}
\end{figure}

We calculated the photon sphere generator satisfying the boundary condition by numerical integration of the null geodesic equation.
We found that (i) the generator is uniquely determined, (ii) asymptotes to $r=3M_1$ from outside in the past direction, and (iii) asymptotes to $r=3M_2$ from inside in the future direction.
The generator is shown in Fig.~\ref{fig:psgenerator-numerical} in $r$-$v$ plane with shadow edge orbits that closely approaches the generator.
Although the Vaidya spacetime with the mass function~\eqref{eq:massfunction-cos} is locally isometric to Schwarzschild spacetimes in the past and future, the photon sphere does not coincide with those of the Schwarzschild spacetimes.
Even if the spacetime is locally static, the photon sphere is not static there.
The shadow edge orbits are obtained by integrating the null geodesic equation in the past direction from each points of the observer's world line at $r=300$.
Each orbit determines the apparent shadow size at each time $v$.
Note that, for the observer at constant radius, time interval of the ingoing null coordinate $v$ is the same as the outgoing coordinate $u$, which is called ``an observer's time" for an observer on $\mathscr{I}^+$.
Figure~\ref{fig:psgenerator-numerical-logplot} shows the shadow edge orbits radius subtracted by the photon sphere generator radius.
We can see that the shadow edge orbits 
asymptote to the photon sphere generator exponentially in time.

\begin{figure}[h]
\includegraphics[width=280pt]{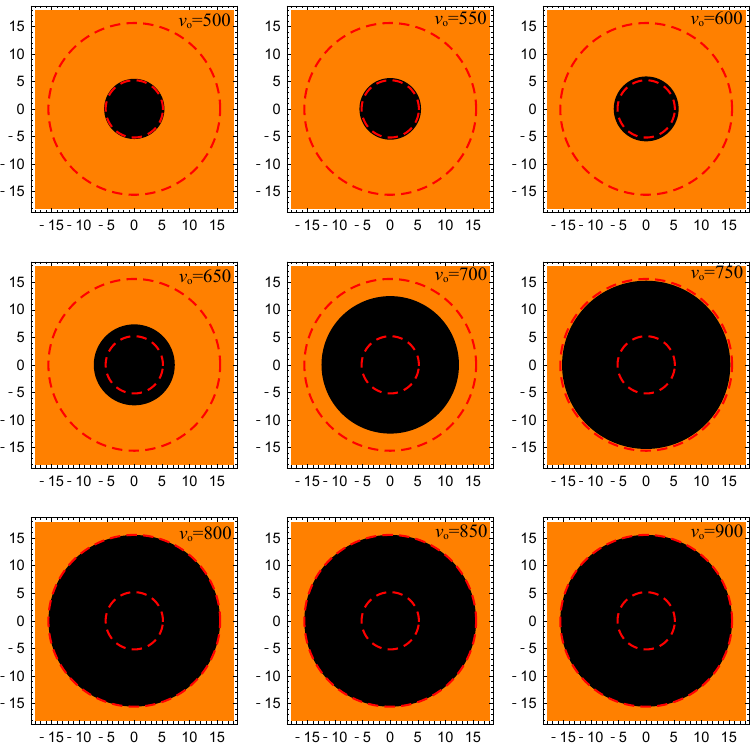}
\caption{\label{fig:shadowimage-numerical} 
Image of the black hole shadow observed at $r=300$ for $v=500, 550, 600, 650, 700, 750, 800, 850,$ and $900$ in the Vaidya spacetime with the mass function~\eqref{eq:massfunction-cos}. 
Parameters are the same as those in Fig.~\ref{fig:psgenerator-numerical}.
The distance from the center corresponds to the impact parameter observed at $r=300$, and the red dashed lines are $b=3 \sqrt{3} M_1$ and $b=3 \sqrt{3}M_2$ for the inner and outer, respectively.
}
\end{figure}

\begin{figure}[h]
\includegraphics[width=200pt]{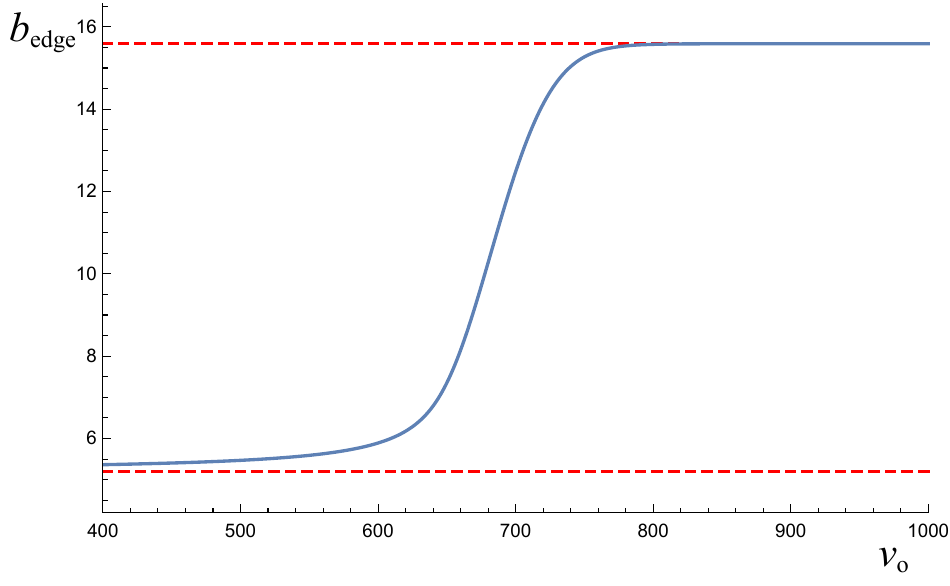}
\caption{\label{fig:vo-bo_graph-cos} 
Time evolution of the shadow edge observed at $r=300$. 
Parameters are same as those in Fig.~\ref{fig:psgenerator-numerical}.
}
\end{figure}

The time evolution of the shadow image is shown in Fig.~\ref{fig:shadowimage-numerical}.
The shadow radius is increasing in time,
and the image is very close to that for the Schwarzschild spacetime with $M=M_1$ at early time and $M=M_2$ at late time, respectively.
Figure~\ref{fig:vo-bo_graph-cos} shows the corresponding time evolution of the shadow edge.

\begin{figure}[h]
\includegraphics[width=170pt]{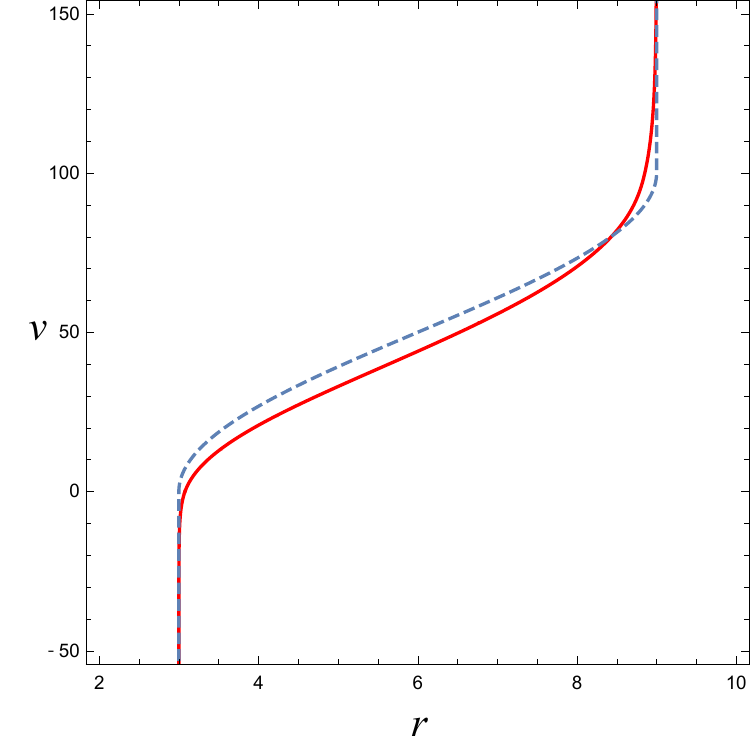}
\caption{\label{fig:psgenerator-numerical2} 
The red and blue dashed line denote the photon sphere orbit and the orbit $r=3m(v)$, respectively. 
The parameters of the spacetime are the same as those in Fig.~\ref{fig:psgenerator-numerical}.
}
\end{figure}

In Fig.~\ref{fig:psgenerator-numerical2},
we plotted the photon sphere orbit and the orbit $r = 3 m(v)$ for comparison.
This shows that
the photon sphere orbit does not coincide with the orbit $r = 3 m(v)$, 
but qualitative behaviors of those two orbits are similar.
As will shown later, for the weakly linear accretion case, the 
deviation of the photon sphere orbit from $r = 3 m(v)$ is 
proportional to the first order of the accretion rate.
\par
\begin{figure}[h]
\includegraphics[width=390pt]{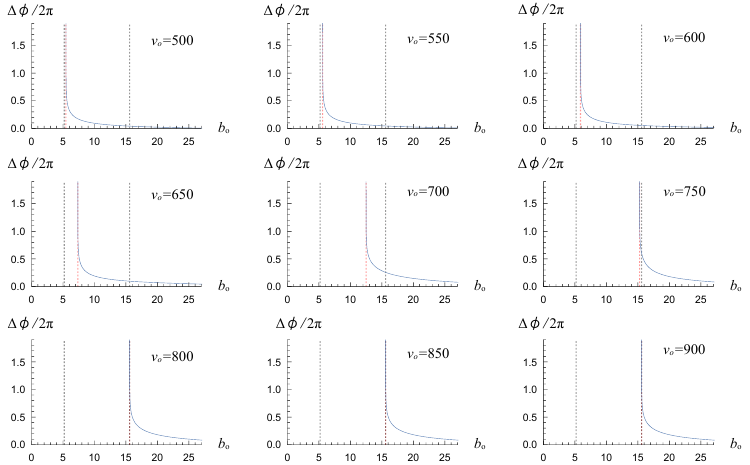}
\caption{\label{fig:deflection_angle-numerical} 
The time evolution of the winding number $n=\Delta \phi/2\pi$ of null geodesics emitted from $r=350$ and observed at $r=300$ as a function of the impact parameter $b_o$. We took the same parameters as in Fig.~\ref{fig:psgenerator-numerical}. The gray dashed lines are $b=3\sqrt{3}M_1$ and $b=3\sqrt{3}M_2$ for the left and right, respectively, and the red dashed line is the impact parameter of the shadow edge orbit $b_{\text{edge}}$ at $v=v_\text{o}$
}
\end{figure}
\begin{figure}[h]
\includegraphics[width=200pt]{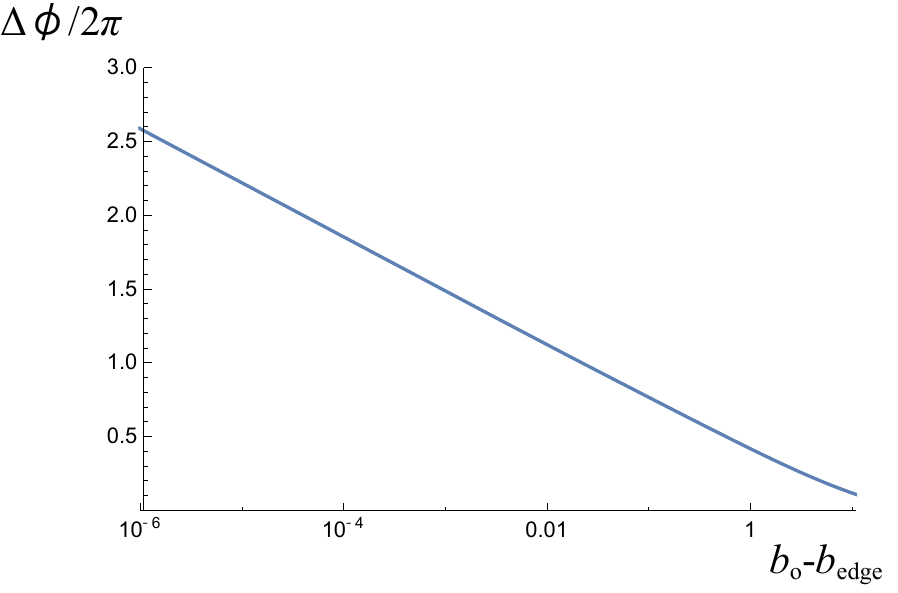}
\caption{\label{fig:deflection_angle-log_plot}
Semilogarithmic scale of Fig.~\ref{fig:deflection_angle-numerical} for $v_\text{o}=700$. 
The horizontal axis is $b_o-b_{\text{edge}}$
and the vertical axis is $\Delta \phi/2\pi$.
}
\end{figure}

We can discuss the winding number $n=\Delta \phi / 2\pi$, 
where $\Delta \phi$ is the total change of $\phi$ subtracted by $\pi$ for a null geodesic which comes from a light source and reaches an observer.
Figure~\ref{fig:deflection_angle-numerical} 
shows the time evolution of the winding number as a function of the impact parameter $b$.
The winding number is divergent at 
the impact parameter $b_{\text{edge}}$ which corresponds to the shadow edge orbit.
In fact, 
similar to the Schwarzschild case, this divergent behavior is also logarithmic.
A typical case is plotted in Fig.~\ref{fig:deflection_angle-log_plot}.
We note that the logarithmic divergent behavior
at the impact parameter which corresponds to the shadow edge orbit 
also can be seen in linear and shell accretion cases which are studied later.

In the following two sections, we investigate the cases where the photon spheres are derived more analytically.
The shadow edge orbits that asymptote to the photon spheres are also shown numerically.

\afterpage{\clearpage}
\newpage

\section{Analytical investigation: linear accretion}
\label{sec:analytical}
\begin{figure}[t]
\includegraphics[width=400pt]{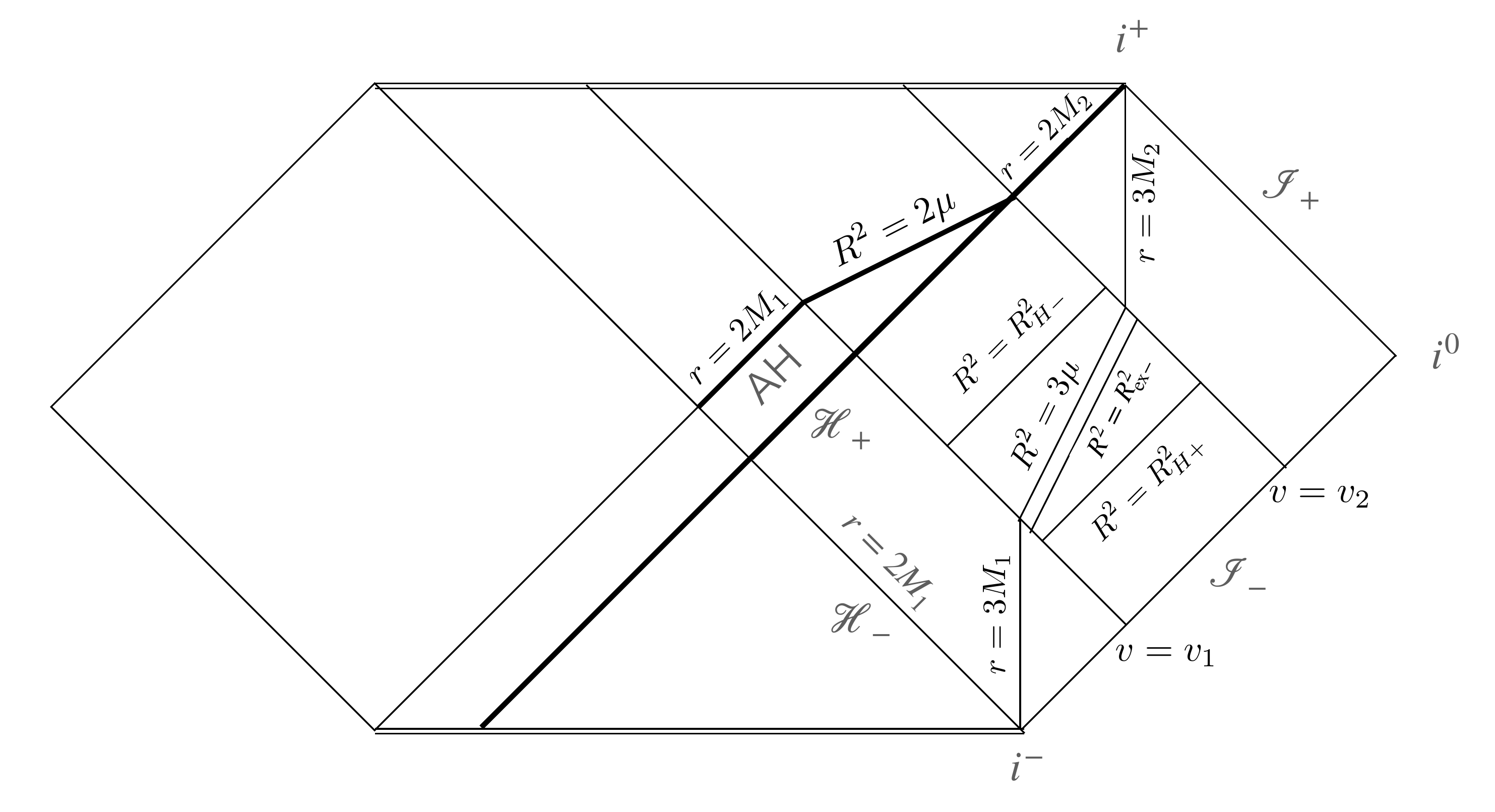}
\caption{\label{fig:causal-structure}
The Penrose diagram of the Vaidya spacetime with the mass function, Eq.~\eqref{eq:massfunction-linear}, for $\mu<1/16$.
The apparent horizon (AH) of Vaidya spacetime is given by $r=2m(v)$~\cite{nielsen}.
The event horizon of the black hole ($\mathscr{H}_+$) is the null hypersurface that matches $r=2M_2$ in the future time domain $v>v_2$.
The analysis of the dynamical domain is given in Sec.~\ref{sec:dynamical_domain}.
}
\end{figure}

Here we consider another case of the Vaidya spacetime,  Eq.~\eqref{eq:vaidyametric}, with the mass function,
\begin{equation}
\label{eq:massfunction-linear}
m(v)=\left\{
\begin{array}{cc}
M_1 & v\le v_1\\
M_1+\mu\left(v-v_1\right) & v_1<v\le v_2\\
M_2=M_1+\mu\left(v_2-v_1\right) & v>v_2
\end{array}
\right. .
\end{equation}
The static time domains correspond to Schwarzschild spacetime with masses $M_1$ and $M_2$.
The mass linearly increases in $v$ in the intermediate dynamical time domain.
We assume $M_1>0$ and $0<\mu<1/16$.
The causal structure of the dynamical region is given as a part of the diagram in Fig.~1 of Ref.~\cite{Hiscock_1982}.
Thus, our spacetime is an eternal black hole and the Penrose diagram is given by Fig.~\ref{fig:causal-structure}.
\par
\begin{figure}[t]
\includegraphics[width=200pt]{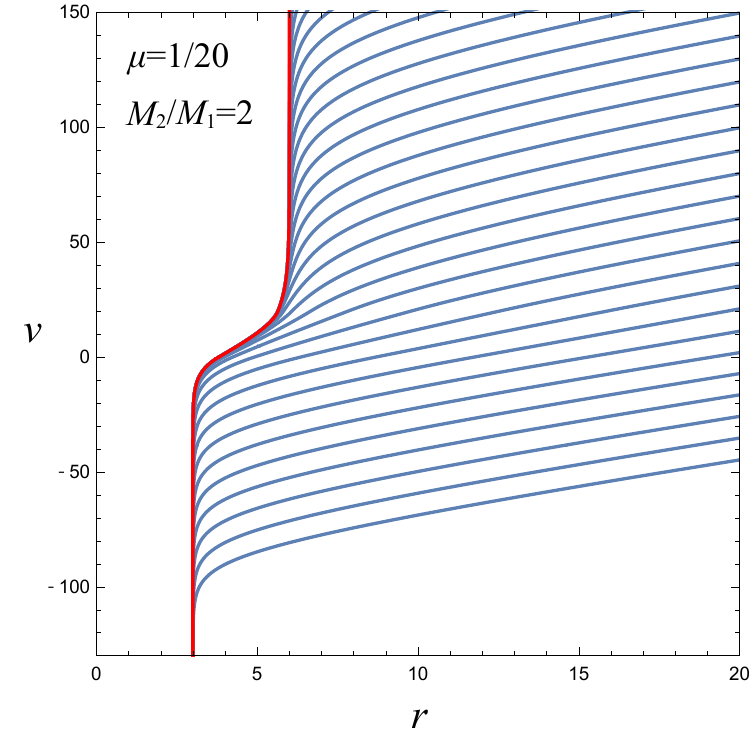}
\caption{\label{fig:psgenerator-linear} 
The orbits of the photon sphere generator (red line) and shadow edge orbits that once approach the generator (blue line) for the Vaidya spacetime with the linear accretion mass function~(\ref{eq:massfunction-linear}).
We set the parameters as $M_1=1$, $M_2=2$, $\mu=1/20$, $v_1=0$, and $v_2=40$.
The photon sphere generator asymptotes to $r=3M_1+0$ in the past and $r=3M_2-0$ in the future.
}
\end{figure}
\begin{figure}[t]
\includegraphics[width=280pt]{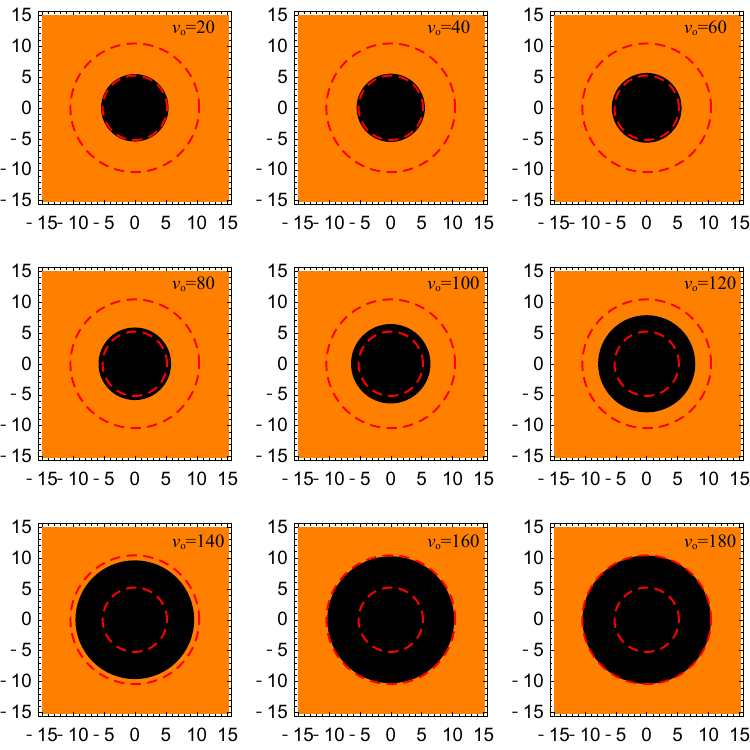}
\caption{\label{fig:Shadowimage-linear} 
Image of the black hole shadow observed at $r=50$ for $v=20, 40, 60, 80, 100, 120, 140, 160$, and $180$ in the Vaidya spacetime with the linear accretion mass function~(\ref{eq:massfunction-linear}). We took the same parameters as in Fig.~\ref{fig:psgenerator-linear}. The distance from the center corresponds to the impact parameter observed at $r=50$, and the red dashed lines are $b=3 \sqrt{3} M_1$ and $b=3 \sqrt{3}M_2$ for the inner and outer, respectively.
}
\end{figure}
\begin{figure}[h]
\includegraphics[width=200pt]{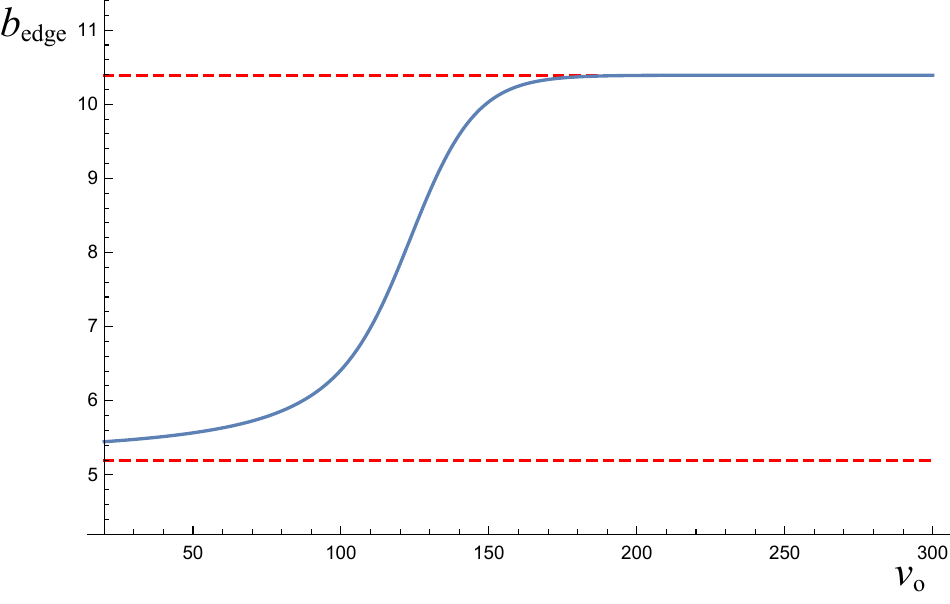}
\caption{\label{fig:vo-bo_graph-self_similar} 
Time evolution of the shadow edge observed at $r=50$. 
We took the same parameters as in Fig.~\ref{fig:psgenerator-linear}.
}
\end{figure}

Because the mass function, Eq.~\eqref{eq:massfunction-linear}, mimics that of Eq.~\eqref{eq:massfunction-cos}, we can expect that the photon sphere of this spacetime also satisfies similar boundary conditions to the ones in the previous section.
Actually, we can see similar behaviors of shadow edge orbits and the photon sphere generator as their limiting surface from numerical integration of the null geodesic equation (Fig.~\ref{fig:psgenerator-linear}). 
The corresponding shadow images and shadow edges are shown in Figs.~\ref{fig:Shadowimage-linear} and~\ref{fig:vo-bo_graph-self_similar}, respectively
(see also Appendix~\ref{sec:linearlargeaccretion} for various accretion rates $\mu$).
As in the previous case, the photon sphere generators asymptote to $r\to 3M_1+0$ and $r\to 3M_2-0$ in the past and future, respectively.
These boundary conditions seem to be generic for this kind of eternal black hole spacetimes.

\par
In the following, we find the photon sphere generator more analytically by assuming that the generators asymptote to $r\to 3M_1+0$ and $r\to 3M_2-0$ in the past and future, respectively.
It is explicitly shown that the deviation of the photon sphere from the hypersurfaces of $r=3M_1$ and $r=3M_2$ depends on the parameters of the geometry, $M_1$, $M_2$, $\mu$, and $v_2-v_1$.

\subsection{Null geodesics in the static domains}
The static time domains, $v\le v_1$ and $v>v_2$, of the Vaidya spacetime are isometric to Schwarzschild spacetimes of masses, $M_1$ and $M_2$, respectively.
We analyze null geodesics in each region in the usual way.
\par
Consider the Hamiltonian for the null geodesic equation,
\begin{equation}
\mathcal{H}=g^{\mu\nu}k_\mu k_\nu =0,
\end{equation}
for the null geodesic tangent $k^\mu=dx^\mu/d\lambda=\dot{x}^\mu$ with the affine parameter $\lambda$.
We assume that the null geodesic lies on the equatorial plane $\theta=\pi/2$ without loss of generality.
Since the basis $\partial_v$ is locally a Killing vector in each static domains and $\partial_\phi$ is globally a Killing vector, we have a locally conserved energy and a globally conserved angular momentum,
\begin{equation}
\label{eq:energy-angular-momentum}
E:=-g(k,\partial_v),\;\;\; L:=g(k,\partial_\phi).
\end{equation}
Then the null geodesic equations of each domain reduces to
\begin{equation}
\label{eq:potential-static}
\dot{r}^2+V_i(E,L;r)=0,\;\;V_i(E,L,r):=L^2f_i(r)r^{-2}-E^2,
\end{equation}
where the functions $f_i(r)=1-(2M_i)/r$ $(i=1,2)$ are the metric component $-g_{vv}$ of each domain.
\par
The effective potentials $V_i$ have maxima at $r=3M_i$.
From the condition $\dot{r}=0$ at $r=3M_i$, we can see that the null geodesics staying the radii have the critical impact parameters $b_{\mr{c}i}^2=27M_i^2$, where an impact parameter is defined by $b^2=L^2/E^2$.
The null geodesics that asymptote to $r=3M_i$ also have the critical impact parameters $b_{\mr{c}i}^2$.
Specifically, the null geodesic that asymptotes to $r=3M_1$ from outside in the past infinity has the critical impact parameter $b_{\mr{c}1}^2$ and satisfies $r>3M_1$ and $\dot{r}>0$ for $v\le v_1$.
The one that asymptotes to $r=3M_2$ from inside in the future infinity has the critical impact parameter $b_{\mr{c}2}^2$ and satisfies $r<3M_2$ and $\dot{r}>0$ for $v> v_2$.

\subsection{Null geodesics in the dynamical domain}
\label{sec:dynamical_domain}
The spacetime locally has a homothetic vector in the dynamical domain~\cite{nielsen,Hiscock_1982}.
Therefore, for $v_1<v\le v_2$, we can take conformally static coordinates by the coordinate transformation $\{v,r\}\to\{T,R\}$,
\begin{equation}
\label{eq:coordtrans-to-TR}
T(v,r)=\ln (v+v_0)-R^*(R(v,r)),\;\; R^2(v,r)=\frac{r}{v+v_0},
\end{equation}
where
\begin{eqnarray}
R^*(R)&:=&\int F^{-1}(R)dR\nonumber\\
&=&-\frac{1}{2(R_{\mr{H}+}^2-R_{\mr{H}-}^2)}\left[R_{\mr{H}+}^2\ln |R_{\mr{H}+}^2-R^2|-R_{\mr{H}-}^2\ln |R^2-R_{\mr{H}-}^2|\right],\nonumber\\
F(R)&:=&\frac{1}{2R}\left(f(R)-2R^2\right)=\frac{1}{2R}\left(1-2\mu R^{-2}-2R^2\right),\nonumber
\end{eqnarray}
where $v_0:=(M_1-\mu v_1)/\mu$, and $f(v,r)=1-2\mu (v+v_0)/r=1-2\mu /R^2=:f(R)$ for $v_1<v\le v_2$.
The time coordinate basis $\partial_T$ is the homothetic vector and $\partial_R$ is the radial basis orthogonal to $\partial_T$.
The metric in the dynamical time domain is then given by
\begin{eqnarray}
\label{eq:conformally-static-metric}
ds^2&=&\Omega^2\left[\frac{2}{R}\left(-FdT^2+F^{-1}dR^2\right)+R^2d\Omega^2\right],\\
\Omega^2&=&(v+v_0)r=e^{2(T+R^*)}R^2.
\end{eqnarray}
The radii $R_{\mr{H}\pm}$ given by
\begin{equation}
\label{eq:RH+-}
R_{\mr{H}\pm}:=\frac{1}{2}\sqrt{1\pm\sqrt{1-16\mu}}
\end{equation}
are the solutions to the equation, $F(R)=0$.
They are the conformal (homothetic) Killing horizons in the sense that $\partial_T$ becomes null.
We can adapt the coordinates $\{T,R\}$ for each of the regions $0<R<R_{\mr{H}-}$, $R_{\mr{H}+}<R<R_{\mr{H}-}$, and $R_{\mr{H}+}<R<\infty$.
However, only the region $R_{\mr{H}+}<R<R_{\mr{H}-}$ is conformally static because $\partial_T$ is timelike there.
The causal structure is investigated in Ref.~\cite{Hiscock_1982} for the maximal extension, corresponding to change of the range $v\in(v_1,v_2)$ to $v\in(v_1-M_1/\mu,\infty)$.
\par
The basis $\partial_T$ is the homothetic vector which is timelike in the conformally static region.
Thus, for \blue{the} null geodesic with the tangent $k^\mu=dx^\mu/d\lambda$ in the dynamical domain, we have a locally conserved ``energy,"
\begin{equation}
\label{eq:def-conformal-energy}
C:=-g(k,\partial_T),
\end{equation}
in addition to the globally conserved angular momentum, $L$.
Then the null geodesic equation reduces to
\begin{equation}
\Omega^4\dot{R}^2+\frac{1}{2}FR^{-1}L^2-\frac{R^2}{4}C^2=0.
\end{equation}
The parameter transformation $\lambda\to\widetilde{\lambda}(\lambda)$ given by
\begin{equation}
\frac{d\lambda}{d\widetilde{\lambda}}=\Omega^2
\end{equation}
further reduces the equation to
\begin{equation}
\label{eq:conformal-pontial-problem}
{R'}^2+U(C,L;R)=0,\;\;U(C,L;R):=\frac{1}{2}FR^{-1}L^2-\frac{R^2}{4}C^2,
\end{equation}
where ${}'=d/d\widetilde{\lambda}$.
\par
\begin{figure}[h]
\centering
\includegraphics[width=200pt]{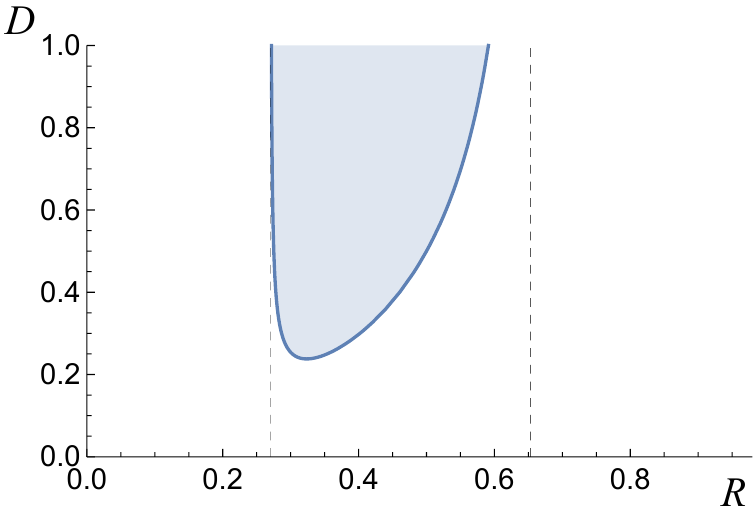}
\caption{
\label{fig:conformal-potential}
$R$-$D$ plane for null geodesics in the dynamical region of the Vaidya spacetime with the mass function, Eq.~\eqref{eq:massfunction-linear}.
The region of $U(D;R)>0$ (shaded region) is the forbidden region.
The vertical dashed lines are $R=R_{\mr{H}-}$ and $R=R_{\mr{H}+}$.
The accretion rate is set to  $\mu=1/32<1/16$ for example.
}
\end{figure}

The behaviors of null geodesics are characterized in terms of the conformal impact parameter, $D:=L/C$, and the rescaled potential, $U(D;R):=C^{-2}U(C,L;R)=U(1,D;R)$.
The null geodesics are given as horizontal lines in $R$-$D$ plane in Fig.~\ref{fig:conformal-potential}, where the forbidden region $U(D;R)>0$ is shown as the shaded region.
The solution of $R$ to $U(D;R)=0$ and $\frac{dU}{dR}(D;R)=0$ is given by
\begin{equation}
R_{\mr{ex}}:=\frac{1}{\sqrt{2}}\sqrt{1- \sqrt{1-12\mu}}.
\end{equation}
The corresponding critical impact parameter is given by
\begin{equation}
D_\text{c}^2=\frac{(1-\sqrt{1-12\mu})(1-\sqrt{1-12\mu}-6\mu)}{2(8\mu-(1-\sqrt{1-12\mu}))}.
\end{equation}
For the critical orbit, the radius $R_{\mr{ex}}$ corresponds to the maximum of the potential, i.e., $\frac{d^2U}{dR^2}(D_\text{c};R_{\mr{ex}})<0$.
As can be seen from Fig.~\ref{fig:conformal-potential}, null geodesics with an impact parameter $D^2<D_\text{c}^2$ is not reflected by the potential.
Another important radius would be $R=\sqrt{3\mu}$, which corresponds to $r=3M_1$ at $v=v_1$ and $r=3M_2$ at $v=v_2$.
In the case $\mu<1/16$, the conformal Killing horizons $R_{\mr{H}\pm}$ exist and these characteristic radii of the dynamical domain satisfy the relations
\begin{equation}
2\mu<R_{\mr{H}-}^2<3\mu<R_{\mr{ex}}^2<R_{\mr{H}+}^2,
\end{equation}
as shown in Fig.~\ref{fig:causal-structure}.

\subsection{Transition among the potential problems}
From the one-dimensional potential problems in each time domain, Eqs.~\eqref{eq:potential-static} and~\eqref{eq:conformal-pontial-problem}, the null geodesic equation of the spacetime is formulated as a piecewise potential problem.
At $v=v_1$ and $v=v_2$, the potential problems are transformed to other ones by the coordinate transformation, $\{v,r\}\leftrightarrow\{T,R\}$.
\par
At $v=v_i\; (i=1,2)$, the energies and radial velocities are transformed as
\begin{eqnarray}
\label{eq:EtoC}
C&=&-g(k,\partial_T)=-g\left(k,\frac{\partial T}{\partial v}\partial_v+\frac{\partial T}{\partial r}\partial_r \right)\nonumber\\
&=&\frac{(v_i+v_0)f_i(v_i,r)-r}{f_i(v_i,r)}E_i-\frac{r}{f_i(v_i,r)}\dot{r},\\
R'
&=&\frac{d\lambda}{d\lambda'}\left(\frac{\partial R}{\partial v}\dot{v}+\frac{\partial R}{\partial r}\dot{r}\right)\nonumber\\
&=&\frac{R}{2C}\left[-\frac{r}{f_i(v_i,r)}E_i+\frac{(v_i+v_0)f_i(v_i,r)-r}{f_i(v_i,r)}\dot{r}\right],
\end{eqnarray}
for $\{v,r\}\rightarrow\{T,R\}$ and
\begin{eqnarray}
E_i&=&-g(k,\partial_v)=-g\left(k,\frac{\partial v}{\partial T}\partial_T+\frac{\partial v}{\partial R}\partial_R \right)\nonumber\\
&=&e^{-(T+R_*(R))}\frac{f(R)C}{f^2(R)-2R^2}\left[f(R)-R^2+2RR'\right],\\
\dot{r}&=&\left(\frac{\partial r}{\partial T}\dot{T}+\frac{\partial r}{\partial R}\frac{d\lambda'}{d\lambda}R'\right)\nonumber\\
&=&-e^{-(T+R_*(R))}\frac{f(R)C}{f^2(R)-2R^2}\left[R^2+\frac{2(f(R)-R^2)}{R}R'\right],
\end{eqnarray}
for $\{T,R\}\rightarrow\{v,r\}$, where the angular momentum $L$ is globally conserved and the coordinate values are transformed according to Eq.~\eqref{eq:coordtrans-to-TR}.
Note that $\dot{r}=\pm V(C,L;,r)$ and $R'=\pm U(C,L;R)$ are obtained more easily from the potentials once the energies are obtained, however, the information of the signs are then lost.
Note also $e^{-(T+R_*(R))}=M_i/\mu$.

\subsection{Photon sphere generator}
\label{sec:ps-generator}
Here we identify for a null geodesic to be the photon sphere generator.
Without loss of generality, we assume that the generator lies on the equatorial plane $\theta=\pi/2$.
First we define {\it a past critical orbit} and {\it a future critical orbit} as null geodesics that asymptote to $r=3M_1$ from outside in the past direction and $r=3M_2$ from inside in the future direction, respectively.
Then we investigate the conditions for them to be connected successfully in the dynamical domain.
\par
A past critical orbit is given as a null geodesic $\gamma_1$ satisfying the condition,
\begin{equation}
\label{eq:gamma1-condition}
r|_{v=v_1}=r_1:=3M_1(1+\epsilon_1),\;\;
\dot{r}|_{v=v_1}=+\sqrt{-V(E_1,L_1;r_1)},\;\;
k_v=-E_1,\;\;
L_1^2=b_{\mr{c}1}^2E_1^2,
\end{equation}
where $\epsilon_1>0$ and the subscripts $1$ represent the quantities of $\gamma_1$.
From Eqs.~\eqref{eq:coordtrans-to-TR} and~\eqref{eq:EtoC}, we have
\begin{eqnarray}
\label{eq:R1-T1}
R^2&=&R_1^2:=3\mu(1+\epsilon_1),\;\;
T=T_1:=\ln\frac{M_1}{\mu}-R^*(R_1),\\
\label{eq:critical-C1}
C_1
&=&\frac{E_1M_1}{\mu}\left[1-9\mu\frac{(1+\epsilon_1)^2}{1+3\epsilon_1}\left(1+\sqrt{1-\frac{1+3\epsilon_1}{(1+\epsilon_1)^3}}\right)\right].
\end{eqnarray}
In the same manner, a future critical orbit is given as a null geodesic $\gamma_2$ satisfying
\begin{equation}
\label{eq:gamma2-condition}
r|_{v=v_2}=r_2:=3M_2(1-\epsilon_2),\;\;
\dot{r}|_{v=v_2}=+\sqrt{-V(E_2,L_2;r_2)},\;\;
k_v=-E_2,\;\;
L_2^2=b_{\mr{c}2}^2E_2^2,
\end{equation}
where $\epsilon_2>0$.
From Eqs.~\eqref{eq:coordtrans-to-TR} and~\eqref{eq:EtoC}, we have
\begin{eqnarray}
\label{eq:R2-T2}
R^2&=&R_2^2:=3\mu(1-\epsilon_2),\;\;
T=T_2:=\ln\frac{M_2}{\mu}-R^*(R_2),\\
\label{eq:critical-C2}
C_2
&=&\frac{E_2M_2}{\mu}\left[1-9\mu\frac{(1-\epsilon_2)^2}{1-3\epsilon_2}\left(1+\sqrt{1-\frac{1-3\epsilon_2}{(1-\epsilon_2)^3}}\right)\right].
\end{eqnarray}
\par
The critical orbits $\gamma_1$ and $\gamma_2$ are successfully connected if they satisfy
\begin{equation}
\label{eq:jc}
\gamma_1^\mu|_{v=v_2}=\gamma_2^\mu|_{v=v_2},\;\;\dot\gamma_1^\mu|_{v=v_2}=\dot\gamma_2^\mu|_{v=v_2}.
\end{equation}
The latter condition is equivalent to the conditions for the conserved quantities,
\begin{equation}
\label{eq:1stjc}
C_1=C_2,\;\;L_1=L_2,
\end{equation}
because each tangent vector has only two independent components due to the fact that the null geodesics are supposed to be on the equatorial plane and satisfy $\mathcal{H}=0$.
From that $L_1=b_{\mathrm{c}1}E_1=3\sqrt{3}M_1E_1$ and $L_2=b_{\mathrm{c}2}E_2=3\sqrt{3}M_2E_2$,
the equation $L_1=L_2$ implies
\begin{equation}
\label{eq:reduced-1stjc-a}
E_1M_1=E_2M_2.
\end{equation}
From Eqs~\eqref{eq:critical-C1},~\eqref{eq:critical-C2}, and~\eqref{eq:reduced-1stjc-a}, the equation $C_1=C_2$ reduces to
\begin{equation}
\label{eq:reduced-1stjc-b}
\frac{(1+\epsilon_1)^2}{1+3\epsilon_1}\left(1+\sqrt{1-\frac{1+3\epsilon_1}{(1+\epsilon_1)^3}}\right)
=\frac{(1-\epsilon_2)^2}{1-3\epsilon_2}\left(1+\sqrt{1-\frac{1-3\epsilon_2}{(1-\epsilon_2)^3}}\right).
\end{equation}
The relation between $\epsilon_1$ and $\epsilon_2$ is independent of any parameters concerning the spacetime itself, $\mu$, $v_1$, $v_2$, $M_1$, and $M_2$.
\par
The former condition in Eq.~\eqref{eq:jc} is explicitly obtained by extending $\gamma_1$ to the time $v=v_2$ from $v=v_1$ by integrating the null geodesic equation.
For simplicity, we assume that, for $v\in (v_1,v_2)$, $\gamma_1$ and $\gamma_2$ are in the conformally static region spanned by the coordinates $\{T,R\}$.
Then the condition is expressed as
\begin{equation}
\label{eq:2ndjc}
T_2-T_1=\int^{\lambda_2}_{\lambda_1}\frac{dT(\lambda)}{d\lambda}d\lambda,
\;\;R_2-R_1=\int^{\lambda_2}_{\lambda_1}\frac{dR(\lambda)}{d\lambda}d\lambda,
\end{equation}
where $T(\lambda)$ and $R(\lambda)$ are the coordinates of $\gamma_1(\lambda)$ and $\lambda_1$ and $\lambda_2$ are the values satisfying $R(\lambda_1)=R_1$ and $R(\lambda_2)=R_2$, respectively.
Further assuming that $\dot{R}(\lambda)<0$ for $\lambda\in[\lambda_1,\lambda_2]$, which is verified in Appendix~\ref{sec:Rdotsign}, the equations are transformed to the form
\begin{equation}
T_2-T_1=\int^{R_2}_{R_1}\frac{dT/d\lambda}{dR/d\lambda}dR=\int^{R_2}_{R_1}\frac{dT/d\widetilde{\lambda}}{dR/d\widetilde{\lambda}}dR.
\end{equation}
Using Eq.~\eqref{eq:coordtrans-to-TR} and the fact that $v_1+v_0=M_1/\mu$ and $v_2+v_0=M_2/\mu$, the left-hand side becomes
\begin{equation}
T_2-T_1=\ln\frac{M_2}{M_1}-R^*(R_2)+R^*(R_1).
\end{equation}
Using Eqs.~\eqref{eq:def-conformal-energy} and~\eqref{eq:conformal-pontial-problem} and the fact $U(C,L;R)=C^2U(1,D;R)$, the right-hand side reduces to
\begin{equation}
\int^{R_2}_{R_1}\frac{dT/d\widetilde{\lambda}}{dR/d\widetilde{\lambda}}dR=-\int^{R_2}_{R_1}\frac{R}{2F(R)}\left[-U(1,D_1;R)\right]^{-1/2}dR,
\end{equation}
where
\begin{equation}
D_1:=\frac{L_1}{C_1}=\frac{3\sqrt{3}M_1E_1}{C_1}
=-3\sqrt{3}\mu f_1(r_1)\left[R_1^2-f_1(r_1)+R_1^2\sqrt{1-b_{\mr{c}1}^2f_1(r_1)r_1^{-2}}\right]^{-1}.
\end{equation}
Then we have
\begin{equation}
\ln\frac{M_2}{M_1}-R^*(R_2)+R^*(R_1)
=-\int^{R_2}_{R_1}\frac{R}{2F(R)}\left[-U(1,D_1;R)\right]^{-1/2}dR.
\end{equation}
Finally, using Eqs.~\eqref{eq:gamma1-condition},~\eqref{eq:R1-T1},~\eqref{eq:gamma2-condition}, and~\eqref{eq:R2-T2},  we obtain the equation
\begin{eqnarray}
\label{eq:reduced-2ndjc}
\ln\frac{M_2}{M_1}
&=&R^*\left(3\mu(1-\epsilon_2)\right)-R^*\left(3\mu(1+\epsilon_1)\right)
+\int^{3\mu(1+\epsilon_1)}_{3\mu(1-\epsilon_2)}\frac{R}{2F(R)}\left[-U(1,D_1;R)\right]^{-1/2}dR,\nonumber\\
D_1&=&3\sqrt{3}\mu \left[1-9\mu\frac{(1+\epsilon_1)^2}{1+3\epsilon_1}\left(1+\sqrt{1-\frac{1+3\epsilon_1}{(1+\epsilon_1)^3}}\right)\right]^{-1},
\end{eqnarray}
as the former condition of Eq.~\eqref{eq:jc}.
Independently of Eq.~\eqref{eq:reduced-1stjc-b}, Eq.~\eqref{eq:reduced-2ndjc} gives the relation between $\epsilon_1$ and $\epsilon_2$ depending on $M_1$, $\mu$, and $v_2-v_1$, where $M_2=M_1+\mu(v_2-v_1)$ is reducible from them.
\par
In summary, if there exist critical orbits $\gamma_1$ and $\gamma_2$ with the parameters $\epsilon_1$ and $\epsilon_2$, respectively, that satisfy Eqs.~\eqref{eq:reduced-1stjc-a},~\eqref{eq:reduced-1stjc-b} and~\eqref{eq:reduced-2ndjc}, they are the PS generator.
Equation~\eqref{eq:reduced-1stjc-a} determines the parameter scaling of $\gamma_2(\lambda)$ relative to $\gamma_1(\lambda)$.
The combination of Eqs.~\eqref{eq:reduced-1stjc-b} and~\eqref{eq:reduced-2ndjc} determines $\epsilon_1$ and $\epsilon_2$ for given $M_1$, $\mu$, and $v_2-v_1$.

\subsection{Results}
\begin{figure}[h]
\centering
\includegraphics[width=200pt]{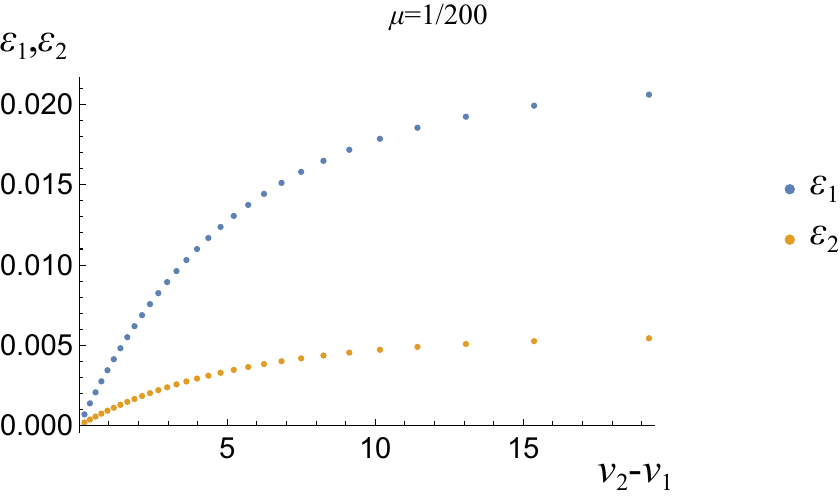}
\includegraphics[width=200pt]{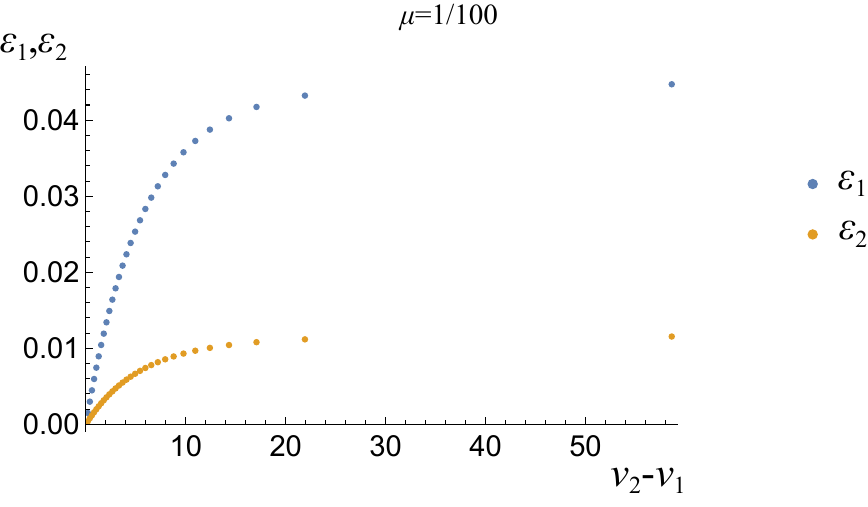}
\includegraphics[width=200pt]{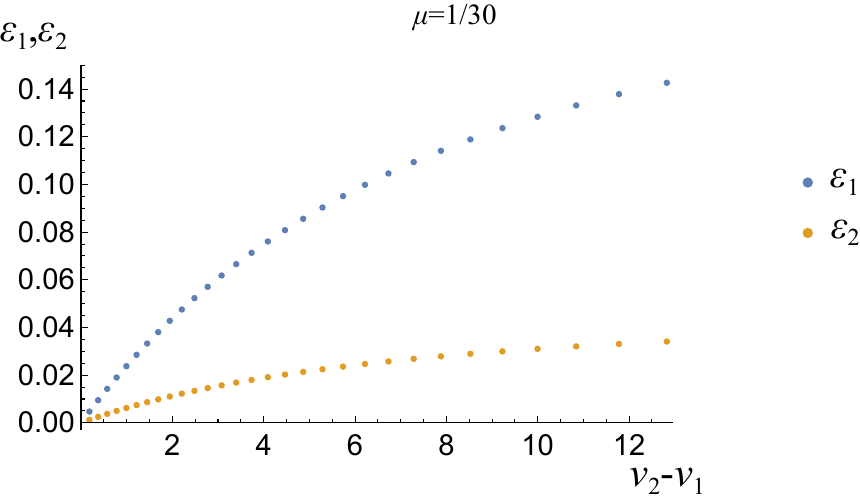}
\includegraphics[width=200pt]{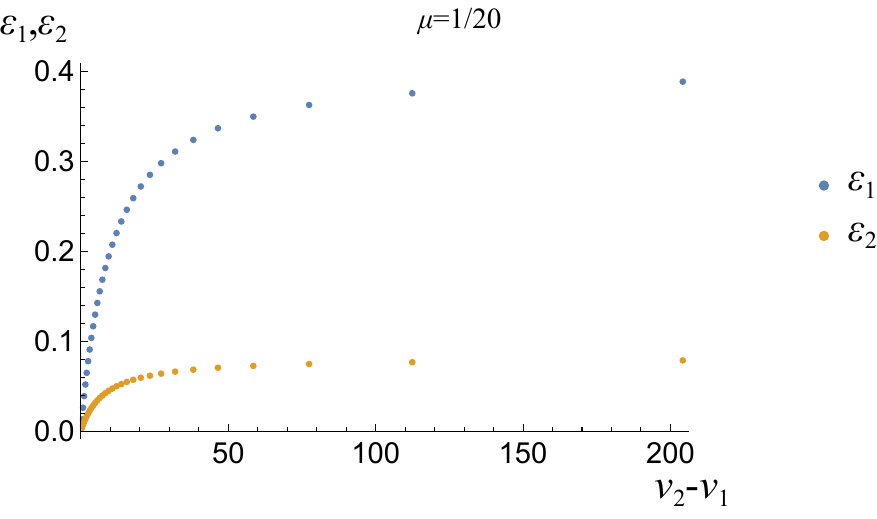}
\caption{
\label{fig:epsilon-results} 
The dots show the numerical results of Eqs.~\eqref{eq:reduced-1stjc-b} and~\eqref{eq:reduced-2ndjc} for varying $v_2-v_1$ with the fixed value of $\mu=1/200, 1/100, 1/30,$ and $1/20$.
The other parameters are chosen so that $M_1=1$, $E_1=1$, and $v_1=0$.
}
\end{figure}

The numerical results of Eqs.~\eqref{eq:reduced-1stjc-b} and~\eqref{eq:reduced-2ndjc} for various values of the parameters, $\mu$ and $v_2-v_1$, are shown in Fig.~\ref{fig:epsilon-results}.
The parameter $M_2$ is given by $M_2=\mu(v_2-v_1)$, and we can take the other parameters as $M_1=1$, $E_1=1$, and $v_1=0$ without loss of generality.
We have investigated the region $\mu<1/18$ because, for $\mu>1/18$, $R_2=\sqrt{3\mu(1-\epsilon_2)}<\sqrt{3\mu}<R_{\mathrm{H}-}$ implying that the past and future critical orbits corresponding to the PS generator are connected outside the conformally static region $R\in(R_{\mathrm{H}-},R_{\mathrm{H}+})$.
This violates the assumption mentioned above Eq.~\eqref{eq:2ndjc}, under which we have derived Eq.~\eqref{eq:reduced-2ndjc}.
\par
\begin{figure}[t]
\includegraphics[width=400pt]{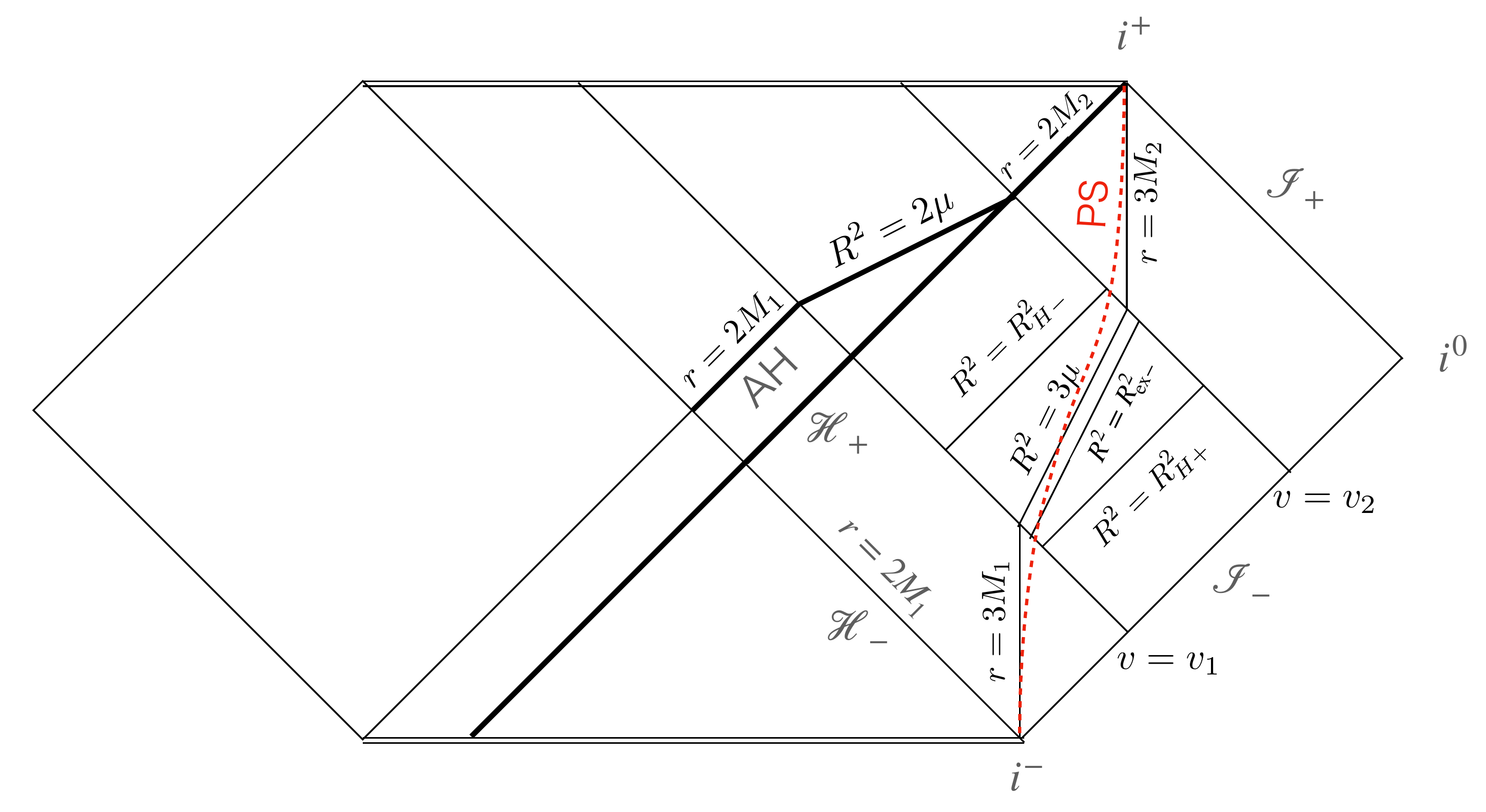}
\caption{\label{fig:ps-penrose}
The dynamical PS (red dashed line) in the temporally accreting Vaidya spacetime.}
\end{figure}

For smaller values, we can analytically determine $\epsilon_1$ and $\epsilon_2$ by linear approximation.
Equation~\eqref{eq:reduced-1stjc-b} is expanded in $\epsilon_1$ and $\epsilon_2$ as
\begin{equation}
-1+9\mu+9\left(\sqrt{3}-1\right)\mu\epsilon_1+\mathcal{O}\left(\epsilon_1^2\right)
=-1+9\mu+9\left(\sqrt{3}+1\right)\mu\epsilon_2+\mathcal{O}\left(\epsilon_2^2\right).
\end{equation}
Equating their orders of magnitude, we have
\begin{equation}
\label{eq:epsilon-ratio}
\epsilon_1=\frac{\sqrt{3}+1}{\sqrt{3}-1}\epsilon_2.
\end{equation}
Equation~\eqref{eq:reduced-2ndjc} is expanded as 
\begin{equation}
\ln\frac{M_2}{M_1}+\frac{9}{2}\frac{\mu^2}{(R_{\mr{H}+}^2-3\mu)(3\mu-R_{\mr{H}-}^2)}({\epsilon}_1+{\epsilon}_2)
=\frac{1-9\mu}{1-18\mu}({\epsilon}_1+{\epsilon}_2)+\mathcal{O}({\epsilon}_1^2,{\epsilon}_2^2).
\end{equation}
For $\ln M_2/M_1=\mathcal{O}({\epsilon}_1,{\epsilon}_2)$, we have
\begin{equation}
{\epsilon}_1+{\epsilon}_2
=\left[\frac{1-9\mu}{1-18\mu}-\frac{9}{2}\frac{\mu^2}{(R_{\mr{H}+}^2-3\mu)(3\mu-R_{\mr{H}-}^2)}\right]^{-1}\ln\frac{M_2}{M_1}
=\ln\frac{M_2}{M_1},
\end{equation}
where we have used Eq.~\eqref{eq:RH+-} in the second equality.
Using Eq.~\eqref{eq:epsilon-ratio}, we finally obtain
\begin{equation}
\label{eq:linear-epsilon}
{\epsilon}_1=\frac{3+\sqrt{3}}{6}\ln\frac{M_2}{M_1},\;\; 
{\epsilon}_2=\frac{3-\sqrt{3}}{6}\ln\frac{M_2}{M_1}.
\end{equation}
The result is valid for the case $\epsilon_1,\;\epsilon_2\ll1$ corresponding to the condition $\delta M/M_1:=(M_2-M_1)/M_1=\mathcal{O}(\epsilon)\ll1$.
Actually, this coincides with $\epsilon_1$ and $\epsilon_2$ for the smaller values of $v_2-v_1$ in Fig.~\ref{fig:epsilon-results}.
In the limit $M_2\to M_1$, the photon sphere coincides with the Schwarzschild photon sphere, $r=3M_1$.
The photon sphere can be described in the Penrose diagram as in Fig.~\ref{fig:ps-penrose}.
\par
From Eqs.~\eqref{eq:R1-T1} and~\eqref{eq:R2-T2}, the radius of the photon sphere is in the range, $R^2\in(3\mu(1-\epsilon_2),3\mu(1+\epsilon_1))$.
The radius $R^2=3\mu$ corresponds to $r=3m(v)$.
Thus, the dynamical photon sphere deviates from but, for weaker accretion, approximately given by three times the Misner-Sharp mass.
In the globally self-similar case of the Vaidya spacetime~\cite{Solanki_2022}, a photon sphere is specified as $R^2=R_{\mr{ex}}^2$, i.e., the maximum of the effective potential $U(C,L;R)$ in Eq.~\eqref{eq:conformal-pontial-problem}.
The photon sphere in our temporarily self-similar case is also different from this case.

\afterpage{\clearpage}
\newpage

\section{Analytical investigation: shell accretion}
\label{sec:analytical-shell}
In the previous section, we mainly focused on the weak accretion case. In this section, as another interesting case which also can be studied analytically, we discuss the null dust thin shell limit.
If we consider the limit of $\mu \to \infty$ and $v_2 - v_1 \to 0$ with $\mu(v_2-v_1) = \delta M = {\rm finite}$ and $v_1=0$, then the mass function $m(v)$ in Eq.~\eqref{eq:massfunction-linear} becomes
\begin{align}
m(v) = M_1 + \delta M \Theta(v),
\label{eq:massfuncshell}
\end{align}
where 
$\delta M = M_2-M_1 (\ge 0)$ and
$\Theta(v)$ is the Heaviside step function
\begin{align}
    \Theta(v) = 
\begin{cases}
  0 & ({\rm for}~v \le 0)
\\ 
  1 & ({\rm for}~v > 0).
\end{cases}
\end{align}
The metric Eq.~\eqref{eq:vaidyametric} with Eq.~\eqref{eq:massfuncshell} describes
the Schwarzschild spacetime with $M_1$ for $v<0$ and $M_2$ for $v>0$, respectively,
and there is a null dust thin shell at $v = 0$.
To study the photon sphere, we discuss the null geodesic on this spacetime.
The tangent of the null geodesic on the equatorial plane 
\begin{align}
k = k^v(\lambda) \partial_v + k^r(\lambda) \partial_r + k^\phi(\lambda) \partial_\phi,
\end{align}
satisfies the geodesic equations $k^\nu \nabla_\nu k^\mu = 0$.
Using $L = g(\partial_\phi, k)$, the $\phi$ component of the geodesic equations can be solved.
The only nontrivial component of the geodesic equations is
\begin{align}
\frac{dk^v}{d\lambda} - \frac{L^2}{r^3} + \frac{(k^v)^2(M_1 + \delta M \Theta(v))}{r^2} =0.
\end{align}
This equation implies that $k^v$ is continuous when the null geodesic goes through the null shell at $v = 0$ surface.\footnote{
If $k^v$ is not continuous, then $dk^v/d\lambda$ contains the Dirac delta function, and then the equation cannot be satisfied.
}
{}From the null condition for $k$
\begin{align}
2 k^r k^v 
+ 
\frac{L^2}{r^2}
+
\frac{(k^v)^2 (2 M_1 - r + 2 \delta M \Theta(v))}{r} = 0,
\end{align}
we obtain the condition for $k^r$ just before and after the null geodesic goes through the null shell
\begin{align}
k^r|_{+0} - k^r|_{-0} = - \frac{\delta M k^v}{r_0},
\label{eq:discontinuity-kr}
\end{align}
where $k^r|_{\pm 0} = \lim_{v \to \pm 0}k^r$ and $r_0$ is the radius of the intersection point of the null geodesic and the null shell.

We wish to find a geodesic which asymptotes to $r=3 M_1$ for $v \to -\infty$ and $r=3 M_2$ for $v \to \infty$, then
the geodesic has the critical impact parameters $b_{\mr{c}1} = 3\sqrt{3}M_1$ for $v < 0$ and $b_{\mr{c}2} = 3\sqrt{3}M_2$ for $v>0$.
Because $L$ is globally conserved,
the relations
\begin{align}
E_1 &= \frac{L}{b_{\mr{c}1}} = \frac{L}{3\sqrt{3}M_1},
\\
E_2 &= \frac{L}{b_{\mr{c}2}} = \frac{L}{3\sqrt{3}M_2},
\end{align}
hold, where $E_1 = -g(\partial_v, k)$ for $v<0$ and $E_2 = -g(\partial_v, k)$ for $v>0$.
{}From the definition of the energy in $v<0$ and $v >0$ regions,
we have
\begin{align}
k^v =
\begin{cases}
\dfrac{r (\sqrt{3}L + 9 M_1 k^r)}{9M_1(r - 2 M_1) } & ({\rm for~~} v <0)
\\ 
\dfrac{r (\sqrt{3}L + 9 M_2 k^r)}{9M_2(r - 2 M_2) } & ({\rm for~~} v >0).
\end{cases}
\label{eq:kvbeforeafter}
\end{align}
The continuity of $k^v$ at the null shell implies
\begin{align}
\frac{\sqrt{3}L + 9 M_1 k^r|_{-0}}{M_1(r_0 - 2 M_1 )} 
= \frac{\sqrt{3}L + 9 M_2 k^r|_{+0}}{M_2(r_0 - 2 M_2)}.
\label{eq:continuity-kv}
\end{align}
From Eqs.~\eqref{eq:discontinuity-kr}, \eqref{eq:kvbeforeafter}
and \eqref{eq:continuity-kv}, we can show
$3 M_1 < r_0 < 3 M_2$, $k^r|_{-0}>0$ and $k^r|_{+0}>0$.\footnote{
Equations~\eqref{eq:discontinuity-kr}, \eqref{eq:kvbeforeafter},
and \eqref{eq:continuity-kv} indicate that
$k^r|_{+0}k^r|_{-0}$ becomes negative if and only if $r_0$ satisfies $3 M_1 +\delta M < r_0 < 3 M_1 + 2 \delta M (< 3 M_2)$.
If $k^r|_{-0}>0$ and $k^r|_{+0}<0$ for $r_0 < 3 M_2$, then the geodesic goes to the black hole horizon.
Thus, we only need to consider the possibility 
$k^r|_{-0}>0, k^r|_{+0}>0$ and $3 M_1 < r_0 < 3 M_2$.
}
Thus, the relations
\begin{align}
k^r =
\begin{cases}
\dfrac{L(r-3 M_1)\sqrt{r + 6 M_1}}{3\sqrt{3}M_1 r^{3/2}} > 0 & ({\rm for~} v <0)
\\ 
\dfrac{L(3 M_2 - r)\sqrt{r + 6 M_2}}{3\sqrt{3}M_2r^{3/2}} > 0 & ({\rm for~} v >0),
\end{cases}
\label{eq:krbeforeafter}
\end{align}
are satisfied.

\begin{figure}[h]
\centering
\includegraphics[width=150pt]{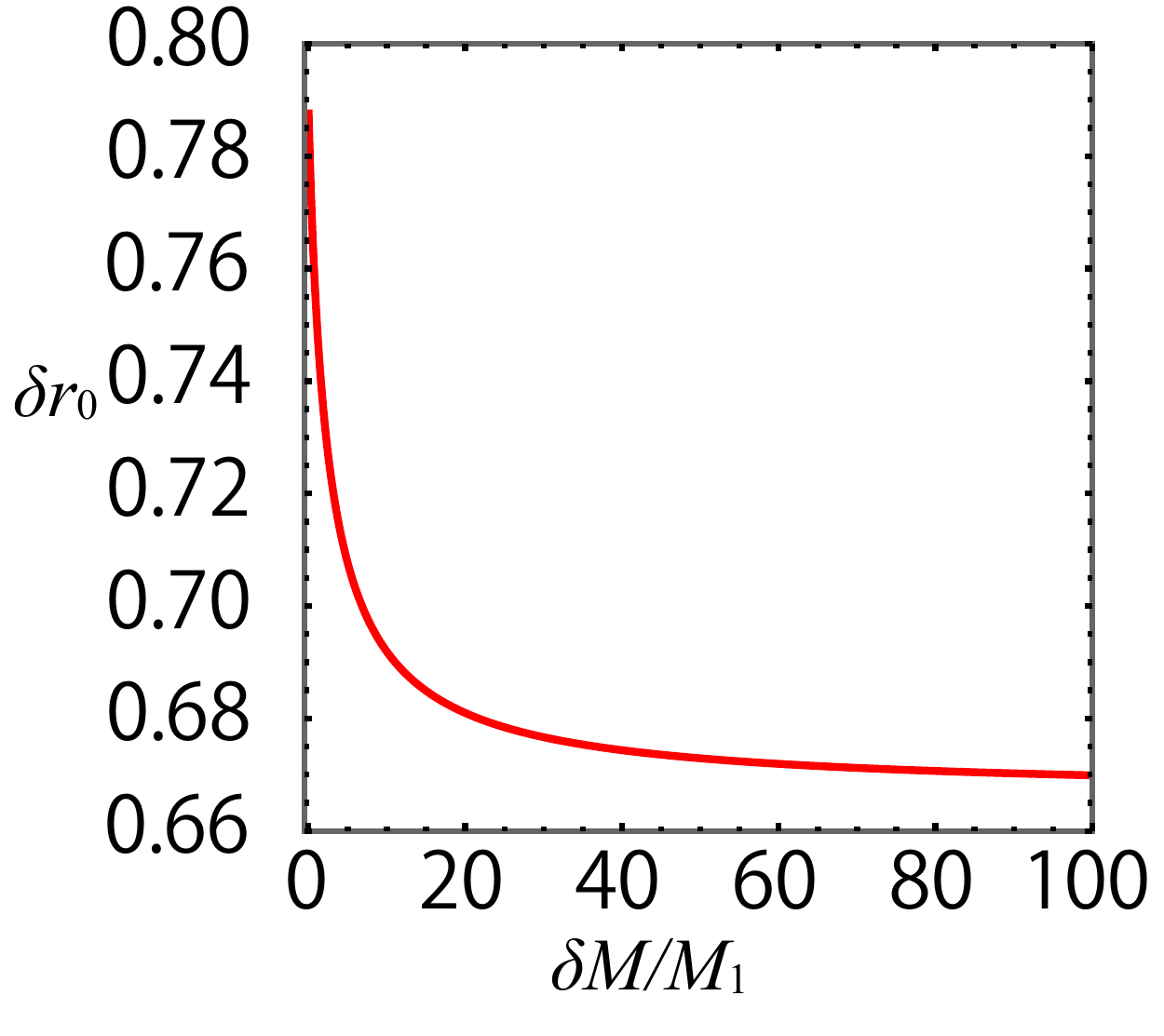}
\caption{
\label{fig:nullshell}
Behavior of $\delta r_0$ as a function of $\delta M$.
}
\end{figure}

\begin{figure}[t]
\includegraphics[width=400pt]{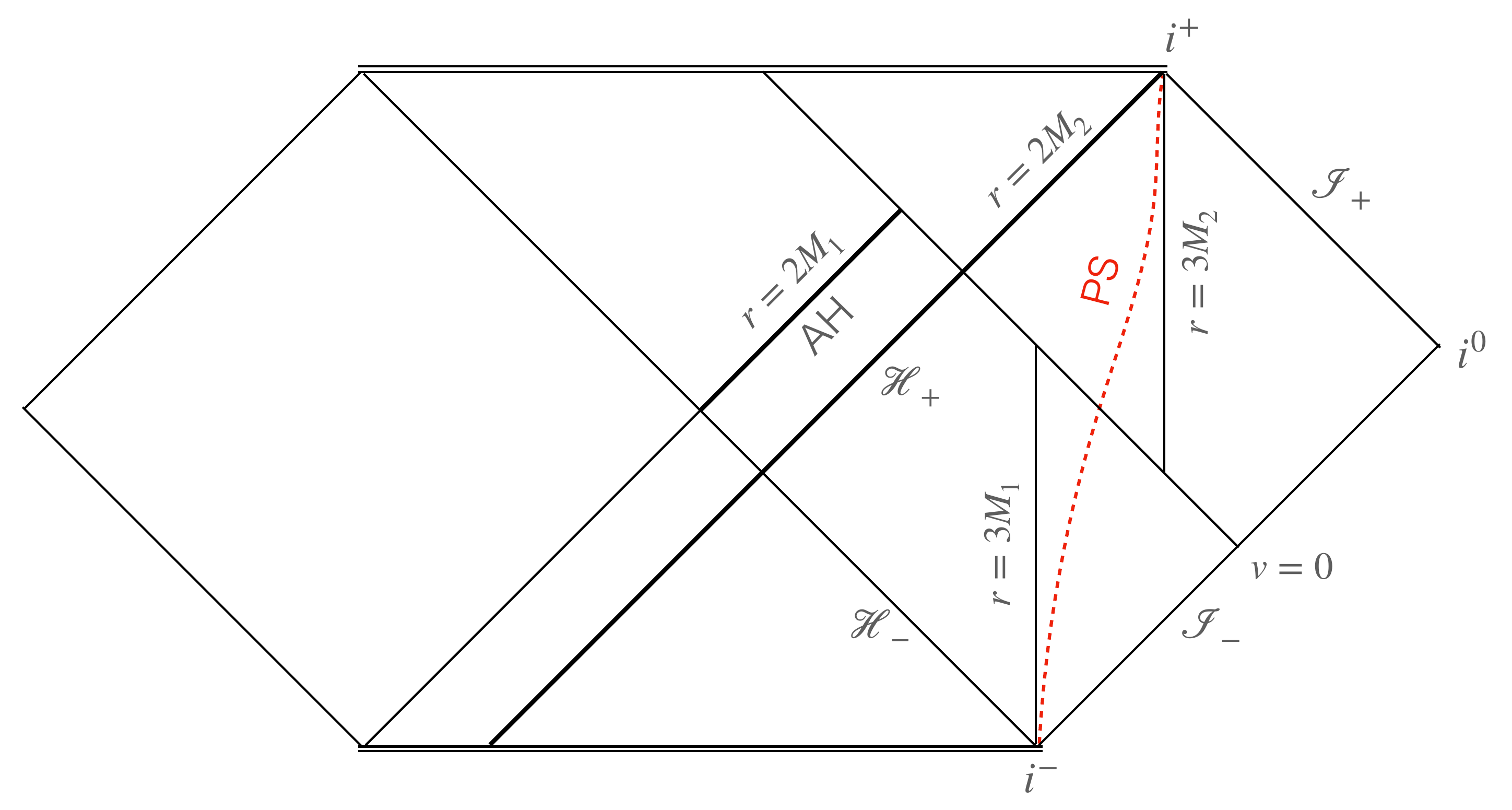}
\caption{\label{fig:ps-penrose-shell}
The dynamical PS (red dashed line) in the Vaidya spacetime with shell accretion.}
\end{figure}

\begin{figure}[t]
\includegraphics[width=200pt]{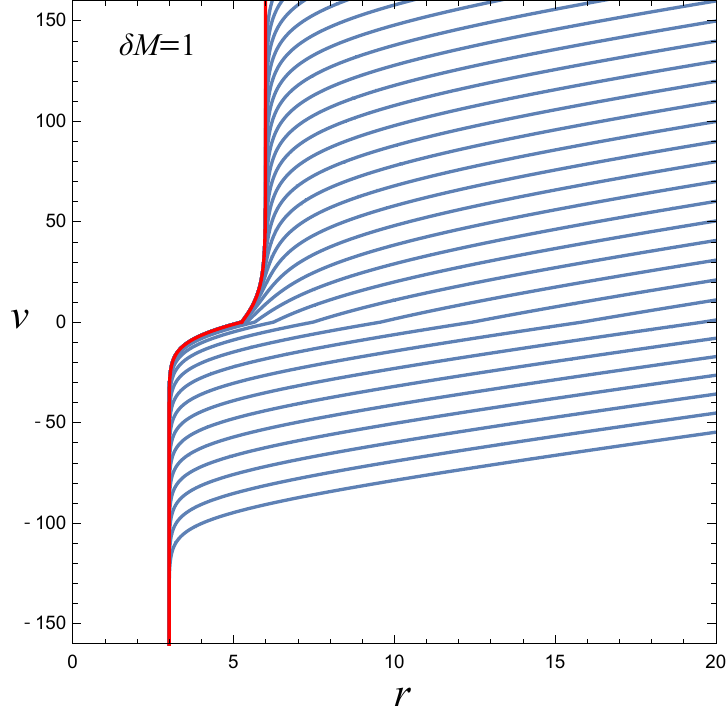}
\caption{\label{fig:psgenerator-nullshell} 
The orbits of the photon sphere generator (red line) and shadow edge orbits that once approaches the generator (blue line)
for the Vaidya spacetime with the shell accretion mass function~\eqref{eq:massfuncshell}.
We took the parameters of the spacetime as $M_1=1$, $M_2=2$.
The photon sphere generator asymptotes to $r=3M_1+0$ in the past and $r=3M_2-0$ in the future.
We can see the discontinuity behavior of $k^r$ in Eq.~\eqref{eq:discontinuity-kr} at $v = 0$.
}
\end{figure}

\begin{figure}[t]
\includegraphics[width=280pt]{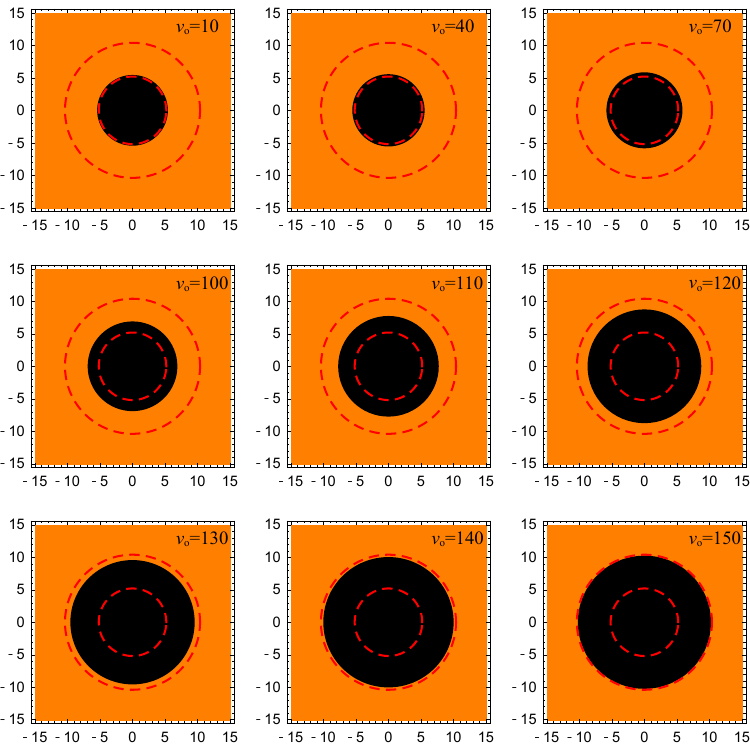}
\caption{\label{fig:Shadowimage-nullshell} 
Image of the black hole shadow observed at $r=50$ for $v=10, 40,70, 100, 110, 120, 130, 140$, and $150$ in the Vaidya spacetime with the shell accretion mass function (\ref{eq:massfuncshell}). We took the same parameters as in Fig.~\ref{fig:psgenerator-nullshell}. The distance from the center corresponds to the impact parameter observed at $r=50$, and the red dashed lines are $b=3 \sqrt{3} M_1$ and $b=3 \sqrt{3}M_2$ for the inner and outer, respectively.
The shadow image for a distant observer is
continuously changes in time even for the shell accretion case.
}
\end{figure}

\begin{figure}[h]
\includegraphics[width=200pt]{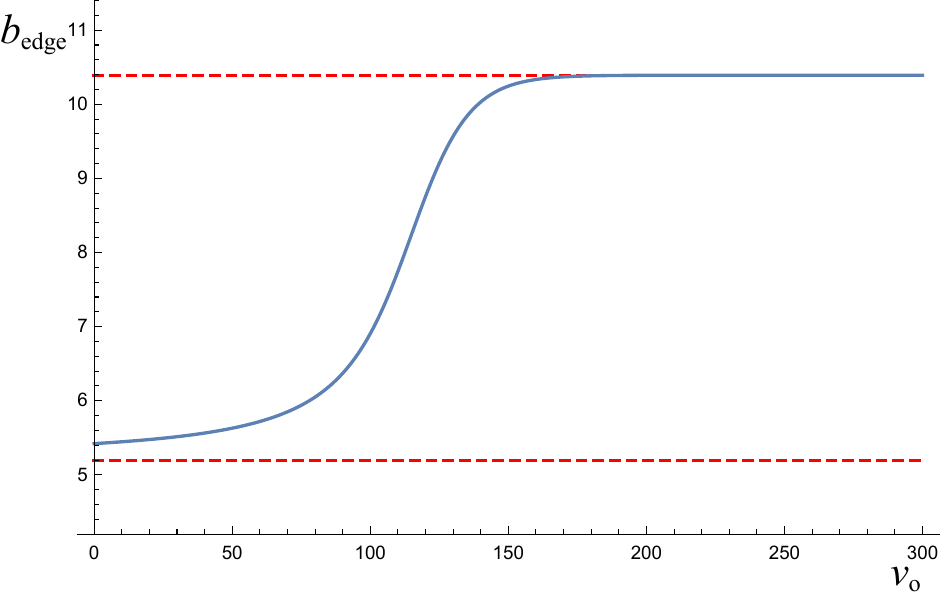}
\includegraphics[width=200pt]{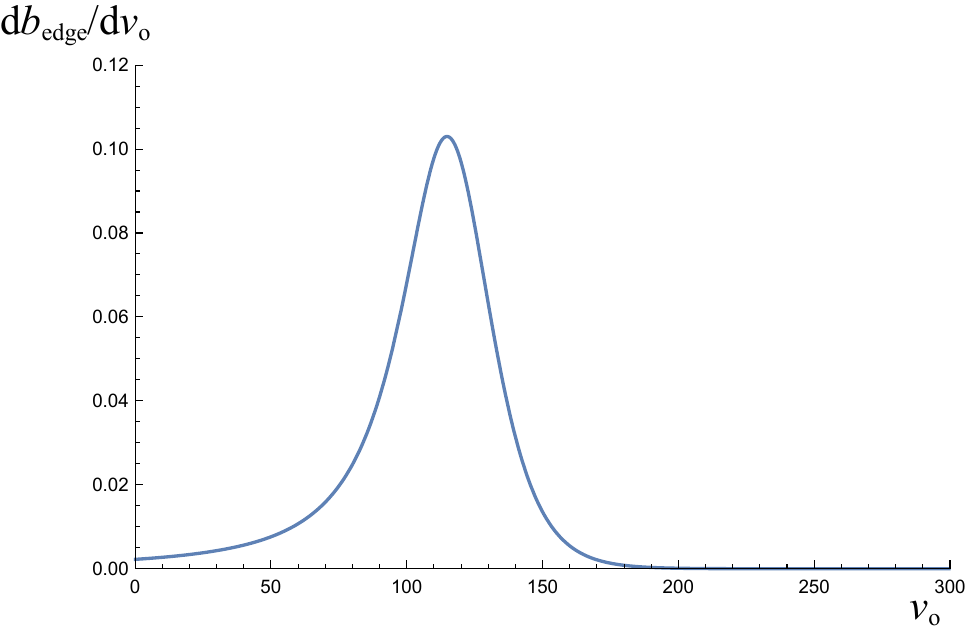}
\caption{\label{fig:vo-bo_graph-shell} 
Time evolution of the shadow edge (left) and the time derivative (right) observed at $r=50$. 
We took the same parameters as in Fig.~\ref{fig:psgenerator-nullshell}.
While the photon sphere generator is continuous but not smooth (see Fig.~\ref{fig:psgenerator-nullshell}),
the shadow edge for a distant observer is continuously and smoothly changing in time.
}
\end{figure}

Equations~\eqref{eq:continuity-kv} and \eqref{eq:krbeforeafter}
determine the value of $r_0$ for the desired null geodesic which corresponds to the photon sphere\footnote{
If we remove the square roots in Eq.~\eqref{eq:eqforr0},
then we obtain a simple equation $r_0^4 - 2(M_1+M_2)r^3 + 27 M_1^2M_2^2=0$.
We should be careful about that the solution of this equation
may not satisfy the original equation~\eqref{eq:eqforr0}.
}
\begin{align}
    \frac{r_0^{3/2}+(r_0-3 M_1)\sqrt{r_0+6 M_1}}{M_1(r_0-2 M_1)}
=
\frac{r_0^{3/2}-(r_0-3 M_2)\sqrt{r_0+6 M_2}}{M_2(r_0-2 M_2)}.
\label{eq:eqforr0}
\end{align}
The solution of Eq.~\eqref{eq:eqforr0} is given by
\begin{align}
r_0 = \frac{1}{2}\left(
(M_1 + M_2)(1 + \alpha_2) + \sqrt{
\frac{2(M_1+M_2)^2(1+\alpha_2)}{\alpha_2} - \frac{3 M_1 M_2 (4+\alpha_1^2)}{\alpha_1}
}
\right),
\label{eq:exactsolr0}
\end{align}
with
\begin{align}
\alpha_1 &=2^{1/3} \left(
\frac{
(M_1+M_2)^2 + (M_2-M_1)\sqrt{M_1^2 + 6 M_1 M_2 + M_2^2}
}{M_1 M_2}
\right)^{1/3},
\\
\alpha_2 &= \sqrt{1 + \frac{3 M_1 M_2(4+\alpha_1^2)}{(M_1 + M_2)^2\alpha_1}}.
\end{align}
We should note that \eqref{eq:exactsolr0} also satisfies Eq.~\eqref{eq:discontinuity-kr}.
To understand the property of Eq.~\eqref{eq:exactsolr0}, it is convenient to introduce
$\delta r_0$ as
\begin{align}
\label{eq:dev-3M1}
    r_0  = 3M_1 \left(1 + \delta r_0 \frac{\delta M}{M_1}\right),
\end{align}
then, $\delta r_0$ represents the deviation of the dynamical photon sphere radius from $3M_1$ at $v=0$.
We note that $\delta r_0$ is a function of $\delta M$ and 
$0 \le \delta r_0 \le 1$ is satisfied.
Equation~\eqref{eq:dev-3M1} also can be written 
as
\begin{align}
\label{eq:dev-3M2}
    r_0=3M_2\left(1-(1-\delta r_0)\frac{\delta M}{M_2}\right),
\end{align}
then $1-\delta r_0$ represents the deviation from $3M_2$ at $v=0$.
If $1 \gg \delta M/M_1$, then $\delta r_0$ approximately behaves as
\begin{align}
\label{eq:delta-r0}
\delta r_0 &=  
\frac{3+\sqrt{3}}{6}
- \frac{1}{18} \left(\frac{\delta M}{M_1}\right)
+ \frac{18+\sqrt{3}}{648} \left(\frac{\delta M}{M_1}\right)^2
- \frac{32+3\sqrt{3}}{1944} \left(\frac{\delta M}{M_1}\right)^3
+{\cal O}(\delta M^4),
\notag\\&\simeq
0.7887
 - 0.05556 \left(\frac{\delta M}{M_1}\right)
+ 0.03045  \left(\frac{\delta M}{M_1}\right)^2
 - 0.01913 \left(\frac{\delta M}{M_1}\right)^3
+{\cal O}(\delta M^4),
\end{align}
If $1 \ll \delta M/M_1$,
$\delta r_0$  approximately behaves
\begin{align}
\delta r_0 &= 
\frac{2}{3}
+
\frac{1}{3}\left(\frac{\delta M}{M_1}\right)^{-1}
-
\frac{9}{8}\left(\frac{\delta M}{M_1}\right)^{-2}
+
\frac{9}{2}\left(\frac{\delta M}{M_1}\right)^{-3}
+ {\cal O}(\delta M^{-4}).
\end{align}
For general cases, the behavior of $\delta r_0$ is plotted in Fig.~\ref{fig:nullshell}.
The dynamical photon sphere whose generator asymptotes to $3M_1+0$ and $3M_2-0$ in the far past and future, respectively, in the Penrose diagram is shown in Fig.~\ref{fig:ps-penrose-shell}.
In Fig.~\ref{fig:psgenerator-nullshell}, 
the photon sphere generator and the null geodesics which asymptotes to it in the past direction are plotted, and 
the discontinuity behavior of $k^r$ in Eq.~\eqref{eq:discontinuity-kr} at $v = 0$ can be seen.
The corresponding shadow images and shadow edges are shown in Figs.~\ref{fig:Shadowimage-nullshell} and~\ref{fig:vo-bo_graph-shell}, respectively.
While the spacetime is suddenly changes at $v = 0$ due to the shell accretion 
and the photon sphere generator is continuous but not smooth (see Fig.~\ref{fig:psgenerator-nullshell}),
the shadow image for a distant observer is continuously and smoothly changes in time.

Finally, we make a remark on the expressions of Eqs.~\eqref{eq:dev-3M1} and~\eqref{eq:dev-3M2}.
Using Eq.~\eqref{eq:delta-r0}, the equations can be written as
\begin{eqnarray}
r_0&=&3M_1(1+\frac{3+\sqrt{3}}{6}\ln\frac{M_2}{M_1})+\mathcal{O}(\delta M^2)=3M_1(1+\epsilon_1)+\mathcal{O}(\delta M^2),\nonumber\\
r_0&=&3M_2(1-\frac{3-\sqrt{3}}{6}\ln\frac{M_2}{M_1})+\mathcal{O}(\delta M^2)=3M_2(1-\epsilon_2)+\mathcal{O}(\delta M^2),
\end{eqnarray}
where we quoted the parameters $\epsilon_1$ and $\epsilon_2$ from the result of the weak accretion limit of the linear accretion case, Eq.~\eqref{eq:linear-epsilon}.
Therefore, although our analysis of the linear accretion case (Sec.~\ref{sec:analytical}) depends on the assumption, $\mu<1/18$, the expressions of the photon sphere radius, $r|_{v=v_1}=3M_1(1+\epsilon_1)$ and $r|_{v=v_2}=3M_2(1-\epsilon_2)$, with Eq.~\eqref{eq:linear-epsilon} are also valid in the shell accretion case corresponding to $\mu\to\infty$.

\afterpage{\clearpage}
\newpage

\section{Relation to photon sphere generalizations}
\label{sec:discussion}
We have specified the photon sphere shaping the black hole shadow.
Our photon sphere should coincide with, be included by, or have some relations to the recently proposed notions generalizing a photon sphere.

\subsection{Photon surface}
In 2001, Claudel, Virbhadra, and Ellis proposed {\it a photon surface} as a geometrical generalization of the Schwarzschild photon sphere~\cite{claudel}:
\begin{definition}
\label{definition:photonsurface}
A photon surface of a spacetime $(M, {g})$ is an immersed, nowhere-spacelike
hypersurface $S$ of $(M, {g})$ such that, for every point $p\in S$ and every null vector ${k}\in T_pS$, there exists a null geodesic $\gamma\colon (-\epsilon,\epsilon) \to M$ of $(M, {g})$ such that $\dot{\gamma}(0) ={k},~ |\gamma|\subset S$.
\end{definition}
The excellent feature of a photon surface is that, in a timelike case, it is a totally umbilic hypersurface~\cite{claudel,perlick}.
The surface is completely characterized by a local geometrical quantity, the extrinsic curvature being pure trace.
See Refs.~\cite{cederbaum,cederbaum_maxwell,yazadjiev_psuniqueness,rogatko_psuniqueness,koga3,tsuchiya,koga:psf-wht,Koga_2021,Kobialko_2021} for the various investigations of photon surface.
\par
In a spherically symmetric spacetime, there are an infinite number of spherically symmetric photon surfaces, or equivalently $SO(3)$-invariant photon surfaces, even in a dynamical case because they are given as solutions to a second order ordinary differential equation~\cite{claudel}.
In other words, any null geodesic with nonzero angular momentum is tangent to some spherically symmetric timelike photon surface.
In this sense, our photon sphere of the Vaidya spacetime is the special one of many photon surfaces that goes to both $i^+$ and $i^-$.
\par
As an application of a photon surface to a black hole shadow, the notion of ``stability" is also important~\cite{koga_2019}.
That is, an unstable photon surface generalizes the usual photon sphere, which is relevant to a black hole shadow, whereas a stable one does the anti-photon sphere, which is irrelevant.
In Appendix.~\ref{sec:psf-stability}, we show our photon sphere is actually an unstable photon surface and therefore, the photon sphere relevant to the black hole shadow.
\subsection{Wandering set}
In the Schwarzschild spacetime,
a null geodesic on the photon sphere is a circular orbit and it comes from the past timelike infinity and go to the future timelike infinity.
This means that although the geodesics on the photon sphere are null, they do not fall into the black hole or escape to the null infinity.
We call such a null geodesic a ``neutral" null geodesic.
Since the generator of the event horizon is also a neutral geodesic,
to exclude it, Siino defines {\it the wandering null geodesic} as follows.
\begin{definition}[\cite{siino_2019,siino_2021}]
A future (past) wandering null geodesic from $p$ is a future (past) complete null geodesic
with infinite number of conjugate points starting from $p$ to the future (past) direction.
A totally wandering null geodesic is a future and past complete null geodesic
with infinite number of conjugate points in both the future and past directions.
\label{wandering_geodesic}
\end{definition}
\noindent
The set of the totally wandering null geodesics is called a {\itshape wandering set}, and it is a generalization of Schwarzschild photon sphere.

For the Vaidya spacetime discussed in this paper, first, according to the Penrose diagram in Fig.~\ref{fig:ps-penrose}, the null generators of the dynamical photon sphere are complete.
Next, let us consider two null generators of the photon sphere starting form a north pole.
We assume that these two geodesics have slightly different azimuth angle.
When one of the geodesics reaches the south pole, the other geodesic also reaches the same point due to the spherical symmetry.
Repeating this argument, we find that 
if these two geodesics intersect infinitely many times, then
there is an infinite number of conjugate points.

To make this intuitive explanation clear, we consider the future directed null geodesic, $k^\mu=dx^\mu/d\lambda$, asymptoting from $r=3M_{1}$ to $r=3M_{2}$ obtained in the previous sections.
We here do not restrict the null geodesic motion to the equatorial plane, $\theta=\pi/2$.
In the future static region, which is described by the Schwarzschild metric,
\begin{align}
  ds_2^2= -f_2(r) dt_2^2 +\frac{dr^2}{f_2(r)}+r^2(d\theta^2+\sin^2 \theta d\phi^2),
\end{align}
where $f_{2}(r)=1-2M_2/r$ and $dt_2= dv-f_2^{-1}(r)dr$,
the conserved energy $E_2$, the conserved angular momentum $L$, and the Carter constant $Q$ are given by
\begin{align}
  E_2=f_2(r)\frac{dt}{d\lambda},
  \quad
  L=r^2 \sin^2 \theta \frac{d\phi}{d\lambda},
  \quad
  Q=r^4 \left(\frac{d\theta}{d\lambda}\right)^2+ L^2 \cot^2 \theta.
  \label{eq:conservation_ene_ang_cater}
\end{align}
From the null condition, we have
\begin{align}
  \frac{d r}{d \lambda}=  E_2 \sqrt{1-\frac{ f_{2}}{r^{2}} \frac{Q+L^2}{E_2^2} }.
\end{align}
Since the null geodesic asymptotes to the spherical photon orbit with $r=3M_{2}$, we have $(Q+L^2)/E_2^2 = 27 M_2^2$ and the polar angle $\theta$ varies in the range $\theta_{\rm min} \le \theta \le \theta_{\rm max}$ where
\begin{align}
  \theta_{\rm min}=\arctan\left( \frac{|L|}{ \sqrt{Q} } \right)
  \quad {\rm and} \quad
  \theta_{\rm max}=\pi -\theta_{\min}.
\end{align}
Hence, the expansion $\tilde{\Theta}$ of the null congruence consisting of nearby null geodesics with the same $E_2$, $L$, and $Q$ is given by
\begin{align}
  \tilde{\Theta}
  =k^{\mu}_{~;\mu}
  =-\frac{ E_2}{r^2} \frac{2r^2 +6M_2r -9M_2^2 }{ \sqrt{r(r+6M_2)}}
  + \epsilon_{\theta} \frac{ 27M_2^2 E_2^2 }{r^2 \sqrt{ Q(\tan^2 \theta - \tan^2 \theta_{\mr{max,min}}) } },
\end{align}
where $\epsilon_{\theta}=\pm 1$ according to the $\theta$ direction of motion.
Since the polar angle $\theta$ repeatedly takes the values $\theta_{\rm min}$ and $\theta_{\rm max}$ for in a finite interval of the affine parameter, the expansion $\tilde{\Theta}$ repeatedly becomes singular.
\footnote{
  The case of $L=0$ corresponds to the intuitive explanation.
}
\footnote{
  Since we choose the null congruence with specific conservation quantities~(\ref{eq:conservation_ene_ang_cater}), the expansion becomes singular at $\theta=\theta_{\rm min},~\theta_{\rm max}$,
  but due to the spherical symmetry, there exists a congruence whose expansion becomes singular at $\theta=\theta_{0},~\pi-\theta_{0}$ for any $\theta_0$.
}
This means that the future directed orbit asymptoting from $r=3M_{1}$ to $r=3M_{2}$ is a future wandering null geodesic.
Note that this conclusion holds for both cases when the geodesics are future and past directed.
Thus, the dynamical photon sphere derived in this paper is a wandering set.

\subsection{Dynamically transversely trapping surfaces}
Yoshino, Izumi, Shiromizu, and Tomikawa introduced the transversely trapping surface in the static and stationary spacetimes as a generalization of the static photon surface by using local quantities~\cite{yoshino_tts}.
Further, they define the dynamically transversely trapping surface (DTTS) as a concept applicable to the dynamical spacetime~\cite{yoshino_dtts}.
The definition of the dynamically transversely trapping surface is given as follows:
\begin{definition}[\cite{yoshino_dtts}]
Suppose $\Sigma$ to be a smooth spacelike hypersurface of a spacetime $\mathcal{M}$.
A closed orientable two-dimensional surface $\sigma_0$ in $\Sigma$ is
a dynamically transversely trapping surface if and only if
there exists a timelike hypersurface $S$ in $\mathcal{M}$
that intersects $\Sigma$ precisely at $\sigma_0$
and satisfies the following three conditions at arbitrary points on $\sigma_0$:
\begin{align}
  \bar{k} &=0,
  \\
  \max \left( \bar{K}_{a b} k^{a} k^{b} \right) & = 0 ,
  \\ ^{(3)}\bar{\mathcal{L}}_{\bar{n}} \bar{k} &\leq 0,
\end{align}
where $\bar{k}$ is the trace of the extrinsic curvature of $\sigma_0$ in the surface $S$, $\bar{K}_{a b}$ is the extrinsic curvature of $S$, $k^a$ are arbitrary future-directed null vectors tangent to $S$, $\bar{n}^a$ is the future-directed unit normal in $S$, and $^{(3)}\bar{\mathcal{L}}_{\bar{n}}$ is a Lie derivative in $S$.
The quantity $^{(3)}\bar{\mathcal{L}}_{\bar{n}}\bar{k}$ is evaluated with a time coordinate in $S$ whose lapse function is constant on $\sigma_0$.
\label{DTTS}
\end{definition}
\noindent
The region in which DTTSs exist is said to be a {\itshape dynamically transversely trapping region}.
If the outer boundary of a dynamically transversely trapping region satisfies the condition $^{(3)}\bar{\mathcal{L}}_{\bar{n}} \bar{k} = 0$, then it is said to be a {\itshape marginally DTTS} and a generalization of Schwarzschild photon sphere.

To discuss whether the dynamical photon sphere in this paper is the marginally DTTS or not, we consider a null geodesic on the equatorial plane in the static regions 
which are described by the Schwarzschild  metric with different masses:
\begin{align}
  ds^{2}_{i}&= -f_{i}(r) d t_i^{2}+\frac{d r^{2}}{f_{i} (r)}+r^{2}(d \theta^{2}+\sin ^{2} \theta d \phi^{2}),
\end{align}
where $f_i(r) = 1-2M_i/r~ (i=1,2)$ and $dt_i = dv - f_i^{-1}dr$.
In this case, we have locally conserved energies and a globally conserved angular momentum Eq.~(\ref{eq:energy-angular-momentum}):
\begin{align}
  E_i &= f_{i}(r) \frac{dt_i}{d\lambda},
  \quad
  L = r^{2} \frac{d\phi}{d\lambda} = E_{i} b_{i}.
\end{align}
From the null condition, we have
\begin{align}
  \frac{d r}{d t_i}=\pm f_{i}(r) \sqrt{1-\frac{b_{i}^{2}}{r^{2}} f_{i}(r)}.
\end{align}
If we obtain a solution of a radial geodesic $r(t_i)$, a photon surface can be constructed due to the spherical symmetry. This photon surface is denoted as $S$.
Then, the induced metric on the photon surface $S$ is given by
\begin{align}
  d s^{2}_{i}=-\alpha_{i}^{2} d t_i^{2}+r^{2}(d \theta^{2}+\sin ^{2} \theta d \phi^{2}),
\end{align}
where the lapse function is given by
\begin{align}
  \alpha_{i}=\frac{b_{i}}{r} f_{i}(r).
\end{align}
We take a spacelike hypersurface $\Sigma_{t_i}$ such that the time coordinate is constant and
the intersection of $\Sigma_{t_i}$ and $S$ is written as $\sigma_{t_i}$ which is a closed two-dimensional surface.
The future-directed unit normal to $\sigma_{t_i}$ in the hypersurface $S$ and the outward spacelike unit normal to $S$ are denoted as $\bar{n}^a$ and $\bar{r}^a$, respectively.
Then, $\bar{k}$ and $ ^{(3)}\bar{\mathcal{L}}_{\bar{n}} \bar{k}$ are given by
\begin{align}
  \bar{k}=\frac{2}{b_{i} f_{i}(r)} \frac{d r}{d t_i}
  \quad
  \text{and}
  \quad
  ^{(3)}\bar{\mathcal{L}}_{\bar{n}} \bar{k}
  =\frac{2}{r^{2}}\left(1-\frac{3 M_{i}}{r}\right).
\end{align}
If we choose the impact parameter of null geodesics from $\sigma_{t_i}$ such that $\bar{k}=0$, i.e. $b_i^2=r^2/f_i(r)$, 
then the first definition of the DTTS is satisfied.
Since the hypersurface $S$ is a photon surface, there is a null geodesic $\gamma\colon (-\epsilon,\epsilon) \to M$ of $(M, {g})$ such that $\dot{\gamma}(0) ={k},~|\gamma|\subset S$. 
Hence, we obtain 
\begin{align}
    \bar{K}_{a b} k^{a} k^{b}=\bar{r}_{a; b} k^{a} k^{b}=(\bar{r}_{a} k^{a})_{;b}k^{b}=0,
\end{align}
along $\gamma$ and then the second definition is also satisfied.
By contrast, whether the third definition is satisfied or not depends on the radius of $\sigma_{t_i}$.
The radius of the dynamical photon sphere in this paper is lager than $3M_1$ in the past region and less than $3M_2$ in the future region.
Hence, the time slice of the dynamical photon sphere is the DTTS in the future region while it is not the DTTS in the past region. 
Since the marginally DTTS is located at $r=3M_2$ in the future region and $r=3M_1$ in the past region, the time slice of the dynamical photon sphere is not the marginally DTTS if a spacelike hypersurface $\Sigma_{t_i}$ is taken such that the time coordinate $t_i$ is constant.

The dynamical photon sphere in this paper clearly depends on the past and future mass, and hence it is determined by a global geometrical structure while the DTTS is defined by local geometrical quantities.
Therefore, as with the relation between the event horizon and the apparent horizon, the dynamical photon sphere in this paper does not necessarily coincide with the DTTS in general.
Note that since the definition of the DTTS depends on a choice of a hypersurface,
there may exist a hypersurface that the dynamical photon sphere in this paper is the DTTS.

\afterpage{\clearpage}
\newpage

\section{Summary and Discussion}
\label{sec:summary}
We have investigated dynamical photon spheres that shape the black hole shadows in the Vaidya spacetime from the causal point of view.
The spacetime has been assumed to be static in the past and future time domains, i.e., isometric to the Schwarzschild spacetime with the mass $M_1$ and $M_2$, respectively.
As a result, we have obtained the photon spheres as hypersurfaces generated by null geodesics that asymptote to $r\to3M_1+0$ and $r\to3M_2-0$ in the past and future, respectively.
Remarkably, the radii of the photon spheres deviate from the Schwarzschild photon spheres $3M_1$ and $3M_2$ even in the static domains.
\par
We have also derived the photon sphere analytically in the case where the evolution of the black hole is linear in the time coordinate $v$ by using the self-similarity of the spacetime there.
The result shows that the photon sphere radius also deviates from the maximum of the conformal effective potential $U(C,L;R)$ as opposed to the entirely self-similar case in Ref.~\cite{Solanki_2022}.
In the weak accretion limit, $\delta M/M_1\ll1$, the deviations of the photon sphere radius from $3M_1$ and $3M_2$ has been derived as $r|_{v=v_1}=3M_1(1+(3+3\sqrt{3})/6\ln M_2/M_1)$ and $r|_{v=v_2}=3M_2(1-(3-3\sqrt{3})/6\ln M_2/M_1)$, respectively.
In the shell accretion limit,
the dynamical photon sphere also locates at the radius between $3M_1$ and $3 M_2$.
Remarkably, in the weak accretion limit of the shell case, i.e. $\mu\to\infty$ and $v_2-v_1\to0$ but $\delta M\ll1$, the expression $r_0=r|_{v=v_1}=3M_1(1+(3+3\sqrt{3})/6\ln M_2/M_1)=r|_{v=v_2}=3M_2(1-(3-3\sqrt{3})/6\ln M_2/M_1)$ holds.
Therefore, we conclude that a dynamical photon sphere shaping a black hole shadow is not determined by local geometry only.
Rather, it depends on global information of the spacetime if one adopts our definitions of a photon sphere and a shadow.
\par
We have discussed the relation between our photon sphere and several notions that generalize a photon sphere.
We have concluded that our photon sphere is a unstable photon surface~\cite{claudel,koga_2019} and a wandering set~\cite{siino_2019,siino_2021}.
Concerning the DTTS~\cite{yoshino_dtts}, we have not found the coincidence with our photon sphere, however, it can be expected from the difference of the viewpoints of the definitions.
\par
It is still challenging to propose, for generic dynamical cases of spacetimes, a generalized definition of a photon sphere as a structure that shapes a black hole shadow.
One of approaches for this problem is to gather many examples in specific cases and to study their essential points.
Then one can check if an existing generalization of a photon sphere is consistent with them or define a new notion so that it is consistent with them.
Our numerical and analytical results would be the good examples in a spherically symmetric spacetime whose dynamics is clearly understood in a physical sense.
As a further investigation, it is important to investigate a dynamical photon sphere in a nonspherically symmetric spacetime.
For example, the photon spheres of the Kastor-Traschen spacetime~\cite{Kastor:1992nn}, a spacetime of two colliding black holes, and its relation to the shadows investigated in Ref.~\cite{Okabayashi_2020, Yumoto:2012kz} are interesting.
\par
Let us apply our results, Eqs.~\eqref{eq:gamma1-condition}, \eqref{eq:gamma2-condition}, and~\eqref{eq:linear-epsilon}, to the observation of M87.
According to Ref.~\cite{Kuo_2014}, the current mass and the accretion rate are estimated as $M_1=3\times 10^9 M_{\odot}$ and $\mu=10^{-3}M_{\odot} \mr{year}^{-1}$.
From $M_2=\mu (v_2-v_1)+M_1$, the radius of the photon sphere after the accretion for the time period of observation, $v_2-v_1$, becomes $r|_{v=v_2}
\simeq 3M_1(1+(3+\sqrt{3})\delta/6)$ where $\delta=\mu(v_2-v_1)/M_1=0.33\times10^{-12}(v_2-v_1)/(1\mr{year})$.
After a few decades, the photon sphere radius evolve only by the ratio $\sim 10^{-11}$.
For black holes with much greater efficient accretion, we might be able to observe the time evolution.

\begin{acknowledgments}
The authors are grateful to T. Harada, T. Ishii, K. Nakao, M. Siino, K. Toma, C. Yoo and H. Yoshino for their fruitful discussions.
%
This work was supported by JSPS KAKENHI Grants No. JP21K20367 (Y.K.), No. JP20H05850 (Y. K.), No. JP20H05853 (Y. K.), NO. JP21J15676 (K.O.), and NO. 20H04746 (M.K.) from the Japan Society for the Promotion of Science.
\end{acknowledgments}

\afterpage{\clearpage}
\newpage

\appendix

\section{Conformal Schwarzschild spacetimes}
\label{sec:conf-sch}
We consider the conformal Schwarzschild spacetime $\Omega^2 g^{\rm Sch}_{\mu \nu}$, where $g^{\rm Sch}_{\mu \nu}$ denotes the Schwarzschild metric, as toy models of the 
black hole spacetimes with an evolving event horizon.
This type of black hole spacetimes were studied in the context of cosmological black holes, e.g., in~\cite{Thakurta1981, Sultana:2005tp}.
Because the light ray orbits are invariant under the conformal transformation,
we can easily discuss the location of the photon sphere.
In particular, we focus on two cases. 
Case I: $\Omega$ is a function of 
the time coordinate $v$  with $dv = dt - (1-2M/r)^{-1}dr$ in the Eddington-Finkelstein coordinates, 
and 
case II: $\Omega$ is a function of 
the time coordinate $\eta$ with $d\eta = dt - 2M dr/(r-2M)$ in the Kerr-Schild coordinates.

\subsection{Case I: Function of the Eddington-Finkelstein time coordinate $v$}
We consider the metric
\begin{align}
ds^2 = \Omega(v)^2\left[
-\left(1 - \frac{2M}{r}\right)dv^2 + 2 dvdr + r^2 (d\theta^2 + \sin^2\theta d\phi^2)
\right].
\end{align}
If the conformal factor is given by
\begin{align}
\Omega^2 = 
\begin{cases}
  1 &({\rm for}~v \le v_0)
\\ 
  1 + \mu (v - v_0) &({\rm for}~v_0 \le v \le v_1)
\\ 
  1 + \mu (v_1 - v_0) := \Omega_{\rm f}^2 & ({\rm for}~v_1 \le v),
\end{cases}
\end{align}
then the spacetime is the Schwarzschild metric with the mass parameter $M$ for $v \le v_0$,
dynamical for $v_0 < v < v_1$, 
and 
the Schwarzschild metric with the mass parameter $\Omega_{\rm f}M$ for $v_1 \le v$.
In this spacetime, the location of the event horizon is $r = 2 M$
and the location of the photon sphere is $ r = 3M$.
Thus, this metric gives a simple toy model of the dynamical black hole spacetime where
we can easily determine the location of the photon sphere.

However, this spacetime has a problematic feature as shown below.
Introducing the new time and radial coordinates $(V,R)$ as $dV = \Omega dv$ and $R = \Omega r$,
the metric becomes
\begin{align}
ds^2 
&= -\left(1 - \frac{2}{R}\left[\Omega M - R^2 \frac{d\Omega /dV}{\Omega }\right]\right)dV^2 + 2dVdR + R^2(d\theta^2 + \sin^2\theta d\phi^2).
\end{align}
This is similar to the form of the Vaidya metric, but 
the Misner-Sharp mass 
\begin{align}
M_{\rm MS} =\Omega M - R^2 \frac{d\Omega /dV}{\Omega }
\end{align}
depends on $R$.
If the horizon area is increasing in time, then the function $d\Omega/dV$ is positive.
Then, the Misner-Sharp mass for $v_0 \le v \le v_1$ becomes negative at the large distance.

\subsection{Case II: Function of the Kerr-Schild time coordinate $\eta$}
Next we consider the metric
\begin{align}
ds^2 = \Omega(\eta)^2\left[
-d\eta^2 + dr^2 +r^2(d\theta^2 + \sin^2\theta d\phi^2)
+
\frac{2M}{r}(d\eta + dr)^2
\right].
\end{align}
If we choose the conformal factor as
\begin{align}
\Omega^2 = 
\begin{cases}
  1 &({\rm for}~\eta \le \eta_0)
\\ 
  1 + \mu (\eta - \eta_0) &({\rm for}~\eta_0 \le \eta \le \eta_1)
\\ 
  1 + \mu (\eta_1 - \eta_0) := \Omega_{\rm f}^2 & ({\rm for}~\eta_1 \le \eta)
\end{cases},
\end{align}
the spacetime has a similar property as the case I.
The Misner-Sharp mass for this spacetime becomes
\begin{align}
M_{\rm MS} = M \Omega + \frac{M r d\Omega/d\eta}{\Omega}(r d\Omega/d\eta - 2 \Omega) + \frac{r^3(d\Omega/d\eta)^2}{2\Omega},
\label{appendix:msmass}
\end{align}
and this is positive everywhere for the above conformal factor.
However, because the last term in Eq.~\eqref{appendix:msmass} 
corresponds to the effect of the expanding Universe,
this spacetime is not appropriate for the toy model of black holes with the localized accreting matter.

\afterpage{\clearpage}
\newpage

\section{Assumption of $\dot{R}<0$}
\label{sec:Rdotsign}
In Sec.~\ref{sec:ps-generator}, to derive Eq.~\eqref{eq:reduced-2ndjc}, we have assumed that the critical orbit $\gamma_1$ satisfies $\dot{R}<0$ for $v\in(v_1,v_2)$.
The assumption is equivalent to that $\gamma_1$ is not reflected by the conformal effective potential $U(C_1,L_1;R)$ for $v\in(v_1,v_2)$.
This is because the case where $\dot{R}>0$ for all $v\in(v_1,v_2)$ is not allowed due to the fact that the radius of $\gamma_1(\lambda)$ for $\lambda\in(\lambda_1,\lambda_2)$ satisfies $R(\lambda_2)=R_2<\sqrt{3\mu}<R_1=R(\lambda_1)$.
Actually, we can see that $\gamma_1$, which is connected to $\gamma_2$, is not reflected by the potential as follows.
\par
\begin{figure}[h]
\includegraphics[width=200pt]{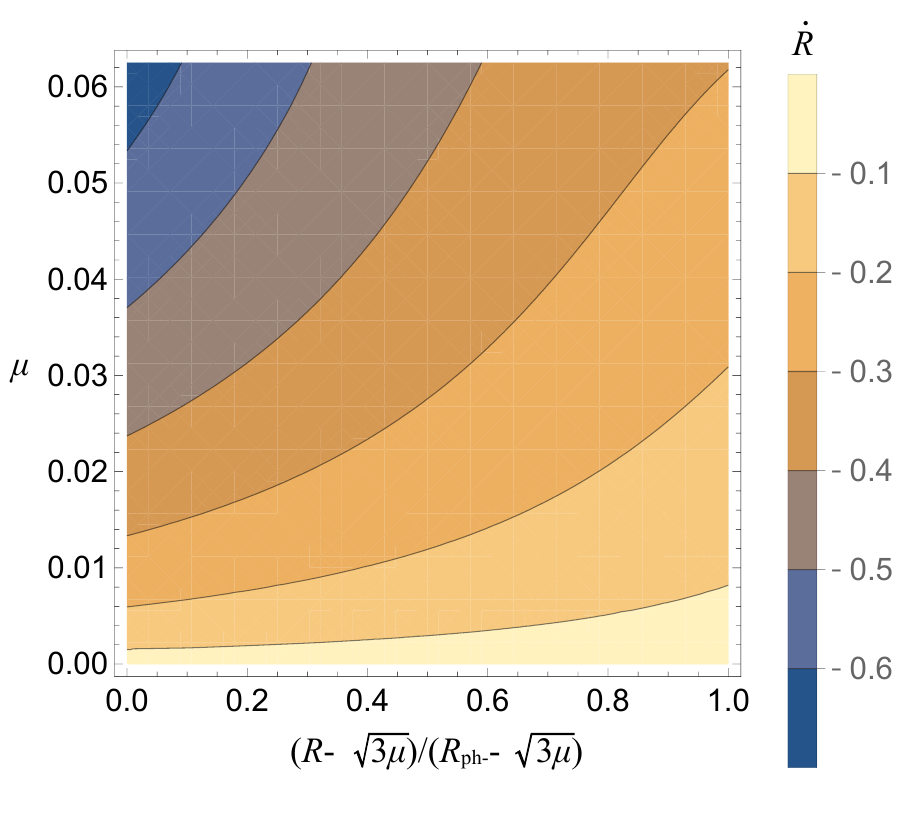}
\caption{
\label{fig:Rdotsign}
The plot of $\dot{R}(R_1)$ for $R_1\in(\sqrt{3\mu},R_{\mr{ph}-})$ and $\mu<1/16$, which is everywhere negative.
}
\end{figure}

First, since $U(C,L;R)$ has only one maximum and no local minimum for $R\in(R_{\mr{H}-},R_{\mr{H}+})$, the orbit $\gamma_1$ can be reflected at most once.
The possible cases for the radius of reflection, $R_{\mr{ref}}$, are $R_{\mr{ref}}<R_2<R_1$ and $R_2<R_1<R_{\mr{ref}}$.
The former case is impossible because it implies that the maximum of the potential, $R=R_{\mr{ph}-}$, is smaller than $R_{\mr{ref}}$.
This contradicts to the fact that $R_2<\sqrt{3\mu}<R_{\mr{ph}-}$.
On the other hand, the latter case implies that $R_2<\sqrt{3\mu}<R_1<R_{\mr{ref}}<R_{\mr{ph}-}$ and initially $\dot{R}>0$ at $R=R_1$.
By the coordinate transformation $\{v,r\}\to\{T,R\}$ given by Eq.~\eqref{eq:coordtrans-to-TR}, we obtain $\dot R=\frac{\partial R}{\partial v}k^v+\frac{\partial R}{\partial r}k^r$ from the null geodesic tangent $k^\mu$ of the photon sphere generator at $v=v_1$ specified by Eq.~\eqref{eq:gamma1-condition}.
We have
\begin{equation}
\dot{R}(R_1)=\frac{E_1}{2R_1M_1(1-2\mu R_1^{-2})}
\left[(1-2\mu R_1^{-2}-R_1^2)R_1^{-3}|R_1^2-3\mu|\sqrt{R_1^2+6\mu}-R_1^2\right]
\end{equation}
as the function of $R_1$.
However, the conditions, $\sqrt{3\mu}<R_1<R_{\mr{ph}-}$ and $\mu<1/16$, lead to $\dot{R}(R_1)<0$ as numerically shown in Fig.~\ref{fig:Rdotsign}.
Therefore, the latter case is also impossible.
\par
As a conclusion, the critical orbit $\gamma_1$ connected to $\gamma_2$, i.e., the null geodesic corresponding to the PS generator, must satisfy $\dot{R}<0$ for $v\in(v_1,v_2)$.

\afterpage{\clearpage}
\newpage

\section{Photon surface stability}
\label{sec:psf-stability}
Our photon sphere derived in Sec.~\ref{sec:analytical} is an $SO(3)$-invariant photon surface~\cite{claudel}.
For the photon surface to be the structure shaping the black hole shadow, it must be ``an unstable photon surface"~\cite{koga_2019}.
\par
Let $S$ and $\gamma$ be our photon surface and a null geodesic along $S$, respectively.
Consider a geodesic deviation $X$ along $\gamma$ that is proportional to the unit normal $n$ to $S$ at $p\in |\gamma|\subset S$.
The null geodesic $\gamma$ is said to be unstable at $p$ if $g(X,\nabla_k\nabla_k X)|_p>0$.
This condition implies that, if a photon orbit along $\gamma$ is parallelly perturbed from $p\in S$ to the direction $n$, then it increases the deviation from $S$ as it propagate in the spacetime.
According to Proposition~1 of Ref.~\cite{koga_2019}, the condition is equivalent to
\begin{equation}
R_{\mu\nu\rho\sigma}k^\mu n^\nu k^\rho n^\sigma<0,
\end{equation}
where $k=\dot\gamma$.
If every null geodesic on $S$ is unstable at every point on $S$, then the photon surface itself is said to be unstable.
For an $SO(3)$-invariant photon surface $S$, if one null geodesic $\gamma$ along $S$ is unstable on every point on $|\gamma|$, $S$ is an unstable photon surface because of the spherical symmetry.
\par
Let $k$ be the null geodesic tangent of the photon sphere generator derived in Sec.~\ref{sec:analytical}.
From Eqs.~\eqref{eq:gamma1-condition} and~\eqref{eq:gamma2-condition}, the tangent is given by
\begin{equation}
k^\mu=\left(E_1f_1^{-1}(r),+E_1V_1(b_{\mr{c}1};r),0,L_1r^{-2}\right)
\end{equation}
and
\begin{equation}
k^\mu=\left(E_2f_2^{-1}(r),+E_2V_2(b_{\mr{c}2};r),0,L_2r^{-2}\right)
\end{equation}
in the static regions $v\leq v_1$ and $v>v_2$, respectively.
The unit normal $n$ to the photon sphere is
\begin{eqnarray}
n_\mu&=&\left(\sqrt{b_{\mr{c}1}^{-2}r^2-f_1(r)},b_{\mr{c}1}^{-1}f_1^{-1}(r),0,0\right),\nonumber\\
n_\mu&=&\left(\sqrt{b_{\mr{c}2}^{-2}r^2-f_2(r)},b_{\mr{c}2}^{-1}f_2^{-1}(r),0,0\right)
\end{eqnarray}
in each region.
Then, we have
\begin{eqnarray}
R_{\mu\nu\rho\sigma}k^\mu n^\nu k^\rho n^\sigma&=&-\frac{3L_1^2M_1}{r}<0,\nonumber\\
R_{\mu\nu\rho\sigma}k^\mu n^\nu k^\rho n^\sigma&=&-\frac{3L_2^2M_2}{r}<0
\end{eqnarray}
in each region.
Therefore, the photon surface is unstable in the static regions.
\par
\begin{figure}[h]
\centering
\includegraphics[width=200pt]{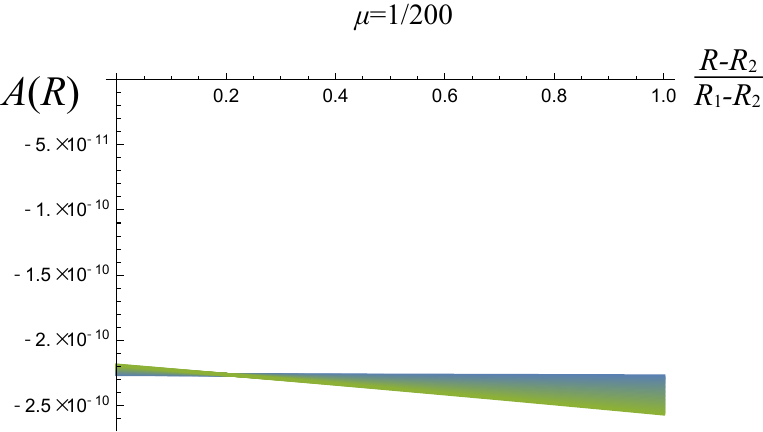}
\includegraphics[width=200pt]{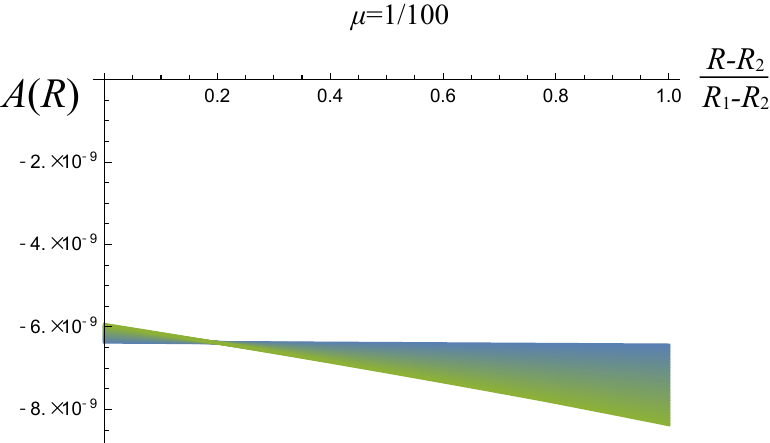}
\includegraphics[width=200pt]{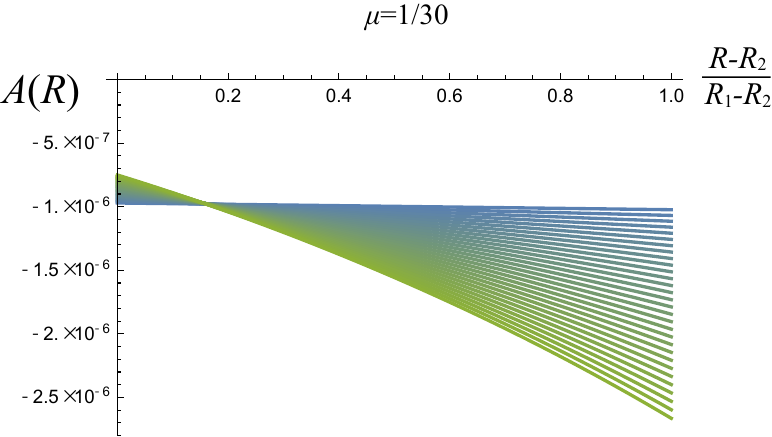}
\includegraphics[width=200pt]{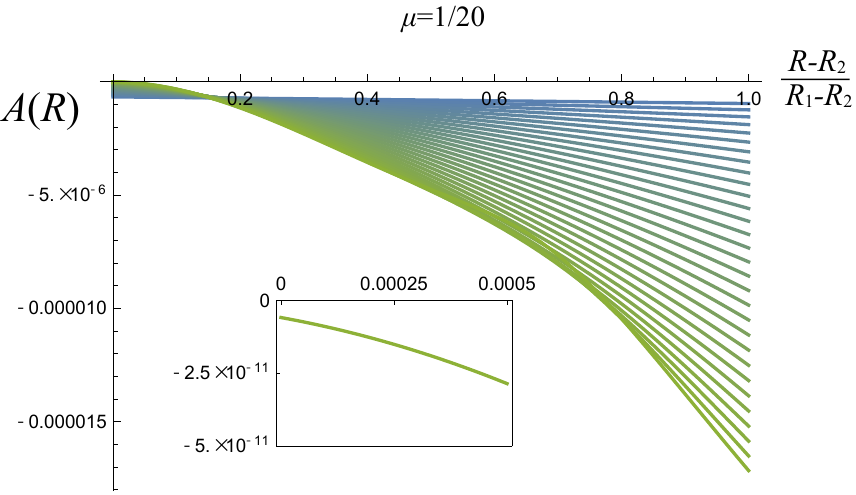}
\caption{
\label{fig:AofR} 
Plot of $A(R)$ for $R\in(R_2,R_1)$.
$\mu=1/200,\;1/100,\;1/30,\;1/20$ are investigated.
For each $\mu$, values of $v_2-v_1$ are chosen as those investigated in Fig.~\ref{fig:epsilon-results}.
The bluer and greener lines correspond to the smaller and greater values of $v_2-v_1$.
We can see $A(R)<0$ for all the ranges.
}
\end{figure}

In the dynamical region, $v_1<v\leq v_2$, the null geodesic tangent is
\begin{equation}
k^\mu=\Omega^{-2}C\left(\frac{R}{2F(R)},-\sqrt{\frac{R^2}{4}-\frac{F(R)}{2R}D^2},0,\frac{D}{R^2}\right).
\end{equation}
The unit normal to the photon sphere is
\begin{equation}
n_\mu=\Omega\left(\sqrt{\frac{2F(R)}{R}},\sqrt{\frac{2}{F(R)R}},0,0\right).
\end{equation}
Then, we have
\begin{equation}
R_{abcd}k^an^bk^cn^d=\frac{1}{4R^{12}F^2(R)}\Omega^{-4}C^2A(R)
\end{equation}
where
\begin{equation}
A(R)=-\mu \left[D^2\left(10R^8-11R^6+(3+22\mu)R^4-12\mu R^2+12\mu^2\right)
-2R^{10}\left(1-\sqrt{1-\frac{2F(R)}{R^3}D^2}\right)\right].
\end{equation}
The conformal impact parameter, $D=D_1$, is given by Eq.~\eqref{eq:reduced-2ndjc} depending on $\epsilon_1$.
$\epsilon_1$ is determined by specifying the parameters $\mu$ and $v_2-v_1$ as derived in the end of Sec.~\ref{sec:analytical}.
The value of $A(R)$ for $R\in(R_2,R_1)$ is plotted in Fig.~\ref{fig:AofR}.
$R_1$ and $R_2$ are determined by $\epsilon_1$ and $\epsilon_2$, respectively, from Eqs.~\eqref{eq:R1-T1} and~\eqref{eq:R2-T2}.
The values of $\mu$ and $v_2-v_1$ are the same as those investigated in Fig.~\ref{fig:epsilon-results} for $\epsilon_1$ and $\epsilon_2$.
The results show that $A(R)<0$ and therefore, the photon surface is also unstable in the dynamical region.

\afterpage{\clearpage}
\newpage

\section{Time evolution of shadow edge for various accretion rates}
\label{sec:linearlargeaccretion}

\begin{figure}[h]
\includegraphics[width=350pt]{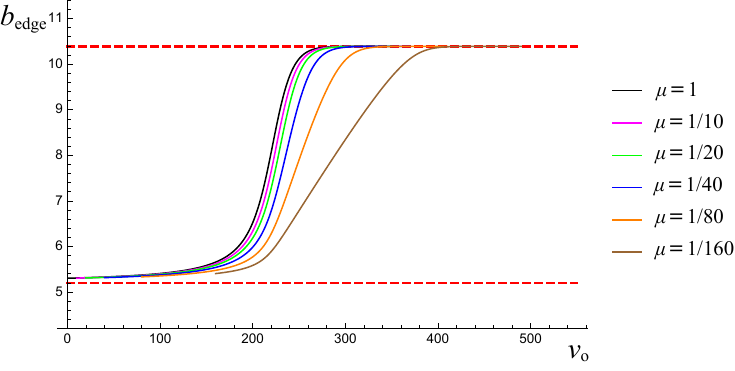}
\caption{\label{fig:linearlargeaccretion1} 
Time evolution of the shadow edge observed at $r = 100$ for various accretion rates $\mu$.
We took the parameters as $M_1 = 1, M_2 = 2, v_1 = 0$.
The lines denote $\mu = 1, 1/10, 1/20, 1/40, 1/80, 1/160$ cases from left to right.
Note that $v_2$ can be read from the equation $M_2 = M_1 + \mu(v_2 - v_1)$.
}
\end{figure}

\begin{figure}[h]
\includegraphics[width=300pt]{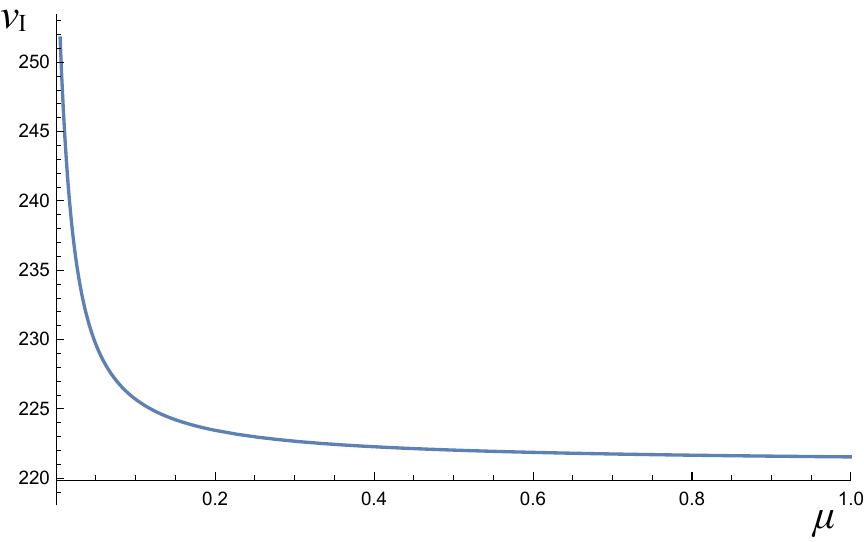}
\caption{\label{fig:linearlargeaccretion2} 
Time at 
the inflection point of $b_{\rm edge}(v_\text{o})$ as a function of $\mu$.
We define $v_\text{I}$ as the time at the inflection point where $d^2b_{\rm edge}/dv_\text{o}^2 = 0$ is satisfied.
We took the parameters as $M_1 = 1, M_2 = 2, v_1 = 0$, and $r_\text{o} = 100$.
}
\end{figure}

\begin{figure}[h]
\includegraphics[width=300pt]{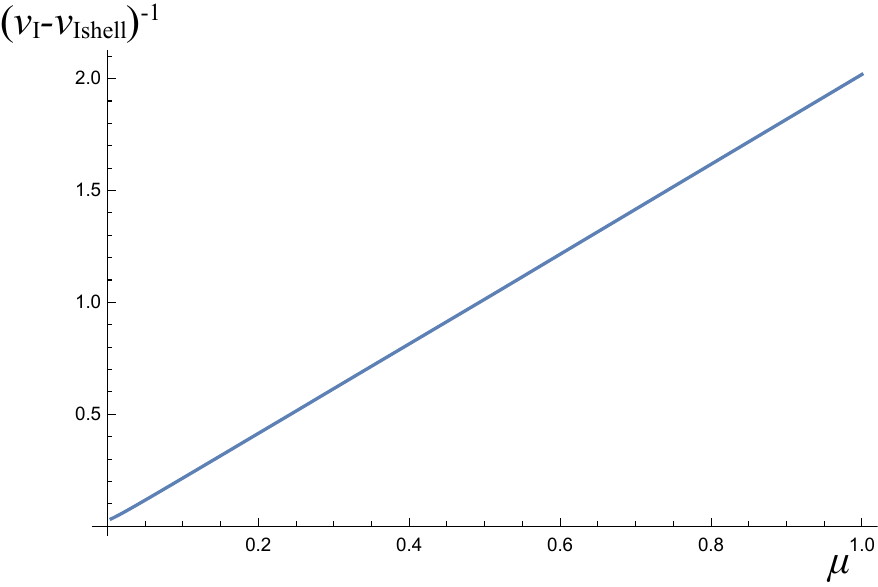}
\caption{\label{fig:linearlargeaccretioninverse} 
Behavior of $v_\text{I}$.
$v_{\text{I} {\rm shell}}$ is defined as $v_\text{I}$ for $\mu \to \infty$ case.
}
\end{figure}

In this section, we study 
the time evolution of shadow edge for the linear accretion model, which is discussed in Sec.~\ref{sec:analytical}, with various accretion rates.
Figure~\ref{fig:linearlargeaccretion1} shows 
the shadow edge observed at $r = 100 M_1$ for various accretion rates $\mu$.
In Fig.~\ref{fig:linearlargeaccretion1}, 
the behaviors of the lines for $\mu = 1$ and $1/10$ are very similar.
This implies that the time evolution of the shadow edge for large accretion rate becomes almost same for a distant observer.
This is the reason why the figures for the linear accretion case in Fig.~\ref{fig:vo-bo_graph-self_similar}
and  for the shell accretion case, which can be considered as the $\mu \to \infty$ limit of the linear accretion case, in Fig.~\ref {fig:vo-bo_graph-shell} are very similar.
In Fig.~\ref{fig:linearlargeaccretion2},
we plot the time at 
the inflection point of $b_{\rm edge}(v_\text{o})$ as a function of $\mu$.
We define $v_\text{I}$ as the time at the inflection point where $d^2b_{\rm edge}/dv_\text{o}^2 = 0$ is satisfied.
We can see that $v_\text{I}$ asymptotes to a value $v_{I{\rm shell}}$ for large $\mu$, where 
$v_{\text{I} {\rm shell}}$ is defined as $v_\text{I}$ for $\mu \to \infty$ case (shell limit).
Figure~\ref{fig:linearlargeaccretioninverse} shows 
$v_\text{I} \sim v_{\text{I} {\rm shell}} + 1/(2 \mu)$ for large $\mu$.

For fixed $M_1$ and $M_2$, if the accretion rate is very large, the spacetime is varying in a short time scale.
Nevertheless, the above results suggest that the shadow edge for a distant observer becomes almost same as the case with not very large accretion.

\afterpage{\clearpage}
\newpage

%
\bibliography{vaidya_shadow}

\end{document}